\documentclass[useAMS,usenatbib]{mn2e}
\usepackage{graphicx}
\usepackage{amsmath}
\usepackage{amssymb}
\usepackage{url}
\usepackage{longtable}
\usepackage{aas_macros}
\usepackage{deluxetable}

\newcommand\ion[2]{#1$\,${\small{#2}}\relax}

\title[SN\,2009ip]{Clues to the Nature of SN\,2009ip II: The Continuing Photometric and Spectroscopic Evolution to 1000 Days}
 
\author[Graham et al.]{M.~L. Graham$^{1,2}$\thanks{E-mail: melissagraham@berkeley.edu}, 
A. Bigley$^{2}$,
J.~C. Mauerhan$^{2}$,
I. Arcavi$^{3,4,5}$,
D.~A. Howell$^{3,5}$, \newauthor
S. Valenti$^{6}$,
C. McCully$^{3,5}$,
A.~V. Filippenko$^{2}$,
and G. Hosseinzadeh$^{3,5}$ \\
$^{1}$ Department of Astronomy, University of Washington, Box 351580, U.W., Seattle, WA 98195-1580 \\
$^{2}$ Department of Astronomy, University of California, Berkeley, CA 94720-3411, USA \\
$^{3}$ Las Cumbres Observatory, Goleta, CA 93117, USA \\
$^{4}$ Einstein Fellow \\
$^{5}$ Physics Department, University of California, Santa Barbara, CA 93106, USA \\
$^{6}$ Department of Physics,  University of California, Davis, 1 Shields Ave, Davis, CA 95616 \\
}

\begin{document}
\pagerange{\pageref{firstpage}--\pageref{lastpage}} \pubyear{2016}

\maketitle

\label{firstpage}

\begin{abstract}

The 2012 brightening of SN\,2009ip was dominated by emission from the interaction of ejecta with the surrounding circumstellar material (CSM) produced by episodic mass loss from the progenitor, complicating the diagnosis of whether the underlying explosion was a true supernova or a nonterminal eruption of a massive star. In this paper, we contribute a time series of optical photometric and spectroscopic observations for SN\,2009ip from 1 to 3 years after the 2012 outburst, collected at the Las Cumbres Observatory and the Keck Observatory. We find that the brightness of SN\,2009ip continues to decline with no deviations from a linear slope of $0.0030\pm0.0005$ $\rm mag\ day^{-1}$ in the $r^{\prime}$ band, and demonstrate that this is similar to both observations and models of CSM-ejecta interaction. We show that the late-time spectra continue to be dominated by the signature features of CSM interaction, and that the large ratio of $L_{\rm H\alpha}/L_{\rm H\beta}\approx40$ implies that the material remains optically thick to Balmer photons (``Case C" recombination). We combine our late-time photometry and spectra with early-time data for SN\,2009ip and provide a comprehensive discussion that incorporates recently published models and observations for transient phenomena dominated by CSM-ejecta interaction, and conclude that the presence of broad H$\alpha$ at early times remains among the best evidence that a terminal supernova has occurred. Finally, we compare our late-time spectra to those of Type\,IIn SN and SN impostors at late phases and find that although SN\,2009ip has some similarities with both types, it has more differences with late-time impostor spectra.
\end{abstract}


\section{Introduction} \label{sec:intro}

The optical transient known as Supernova (SN) 2009ip was discovered in host galaxy NGC 7259 by the Chilean Automatic Supernova Search (CHASE; \citealt{2009CBET.1928....1M}). Archival {\it Hubble Space Telescope (HST)} images indicated that its progenitor star was very massive, $M \approx 50$--80 $\rm M_{\odot}$ \citep{2010AJ....139.1451S}, and subsequent photometry and spectroscopy confirmed that SN\,2009ip was an erupting luminous blue variable (LBV) star \citep{2009ATel.2183....1M, 2009ATel.2184....1B, 2009ATel.2212....1L, 2010ATel.2897....1D, 2011ApJ...732...32F}. In mid-September 2012, SN\,2009ip deviated from its typical variability and exhibited both an increase in brightness and spectral lines with P-Cygni profiles, which indicated the release of material at velocities typical for supernovae \citep[SNe;][]{2012ATel.4412....1S}. The spectrum also exhibited relatively narrow H$\alpha$ emission lines indicative of ejecta interacting with circumstellar material (CSM), as is commonly seen in both Type IIn SNe and ``SN impostors" (e.g., \citealt{2000PASP..112.1532V}). Observations of ejecta interacting with CSM are useful for deducing the mass-loss history of a progenitor star during the late stages of stellar evolution (e.g., as done for SN\,2009ip by \citealt{2013ApJ...768...47O}), but a drawback of CSM interaction emission is that it can overwhelm the flux from the underlying event, making it difficult to identify whether the type of explosion powering the ejecta was a true SN or a massive eruption.

Owing to its ambiguous nature, SN\,2009ip is the subject of many studies and interpretations. High-cadence time-series photometry of SN\,2009ip's 2012 eruption presented by \cite{2013ApJ...763L..27P} shows a $\sim20$ day precursor outburst reaching $I\approx-15$\,mag (the ``2012-A" event) followed immediately by an outburst reaching $I\approx-18$\,mag (the ``2012-B" event). Although seemingly rare, the double-peaked light curve of SN\,2009ip is seen in some typical Type IIn SNe \citep{2014MNRAS.438.1191S,2014ApJ...789..104O}. The question of whether SN\,2009ip was a true SN was first addressed by three papers: \cite{2013MNRAS.430.1801M}, \cite{2013MNRAS.433.1312F}, and \cite{2013ApJ...767....1P}. Mauerhan et al. (2013) suggest that the material velocities indicated by the P-Cygni absorption features and the broad components of Balmer emission lines during the 2012-A event, before the onset of CSM interaction, point to a true SN explosion during an LBV outburst. In contrast, Fraser et al. (2013) argue that there is insufficient evidence of a true SN, pointing out that no nebular emission from nucleosynthetic material is seen, and that the maximum mass of synthesised $^{56}$Ni is $M_{\rm Ni} < 0.02$ $\rm M_{\odot}$ (i.e., quite low). Pastorello et al. (2013) present the full light curve of SN\,2009ip from its LBV outbursts in 2009 through the 2012 event, as well as spectra from its 2011 (nonterminal) outburst in which very high velocity material was also seen in absorption (13,000 $\rm km\ s^{-1}$). They suggest that this supports a nonterminal explanation for the 2012 outburst; however, since it takes relatively little mass to make an absorption feature, this is not unambiguous evidence for or against a true SN.

Further details regarding the nature of SN\,2009ip were discussed by \cite{2014AJ....147...23L}, who present an analysis of the Balmer decrement as evidence for a high-density, thin-disk geometry for the CSM; \cite{2014ApJ...780...21M}, who show that their high-cadence, panchromatic observations illuminate the episodic explosive mass loss of the progenitor star; \cite{2014MNRAS.438.1191S}, who demonstrate that the late-time spectra of SN\,2009ip are most similar to those of core-collapse SNe (e.g., Sn\,2010mc and SN\,1987A, also interpreted as SNe arising from blue supergiant stars with CSM) and do not resemble a return to an LBV-like state, and that the width and persistence of the P-Cygni profile are possible only with a kinetic energy of $>10^{51}$ $\rm erg$, strongly supporting a terminal explosion as the underlying event; \cite{2014ApJ...787..163G}, the prequel to this work, who link features of CSM interaction in the light curve with previously observed eruptions and find that the photometric and spectroscopic evolution to late times supports a SN\,IIn-like explosion; and \cite{2014MNRAS.442.1166M}, whose unique spectropolarimetric time-series observations reveal that the two aspheric components from the explosion and CSM-impact have orthogonal axes of symmetry, and furthermore that this significant asymmetry indicates a total explosion energy of $\sim 10^{51}$ $\rm erg$, thereby establishing a terminal SN explosion as the most plausible physical explanation for SN\,2009ip.

The late-time light curve of SN\,2009ip has been documented by \cite{2015ApJ...803L..26M}, who interpret the ``2012-A" and ``2012-B" events as interactions with two distinct shells of CSM (as opposed to a SN followed by CSM interaction), and conclude that a normal SN is not necessary to explain the total energy released. However, \cite{2015ApJ...803L..26M} assumed spherical symmetry for the explosion and the CSM, which is inconsistent with the spectropolarimetric results of \cite{2014MNRAS.442.1166M}. \cite{2015MNRAS.453.3886F} collate late-time optical to mid-infrared spectra and photometry from 2013 and 2014, showing a continuation of the smooth decline in brightness and minimal changes in the spectral features, which continue to be dominated by the signature of CSM interaction. In late 2015, Thoene et al. (2015) announced that SN\,2009ip had finally faded past its previous faintest observed magnitude, and \cite{2016arXiv160701056S} confirm this with multiband $HST$ photometry from May 2015. They furthermore show that the environment of SN\,2009ip does not appear to contain a significant star-forming region (although they note that such imaging cannot rule out a small one in the immediate vicinity until SN\,2009ip fades). \cite{2016arXiv160701056S} suggest that since young, massive stars are unlikely to be isolated, their observations support a binary mechanism as the cause of SN\,2009ip's variability.

In this paper, we build on our previous work in \cite{2014ApJ...787..163G} and present optical photometric and spectroscopic monitoring of SN\,2009ip from 260 through 1026 days past the start of the ``2012-B" event (2012 Sep. 23 UT). We describe the acquisition and reduction of our observations in \S~\ref{sec:obs}. In \S~\ref{sec:ana} we analyze our data, describing the late-time photometric and spectroscopic qualities and their evolution and providing comparisons with late-time observations and models for transients which, like SN\,2009ip, are dominated by CSM interaction. In \S~\ref{ssec:models} and \ref{ssec:nature} we expand our analysis to reconsider the early-time data of SN\,2009ip in context with recently published models and data for similar transient phenomena. Section~\ref{sec:conc} provides a summary and our conclusions.

\section{Observations}\label{sec:obs}

\begin{table}
\begin{center}
\label{tab:phot}
\begin{tabular}{cccc}
\hline
\hline
\multicolumn{2}{c}{LCOGT $g$ band} & \multicolumn{2}{c}{LCOGT $r$ band}\\
MJD & Magnitude & MJD & Magnitude \\
\hline
56445 & $19.73\pm0.33$  & 56445 & $18.42\pm0.14$ \\
56447 & $19.30\pm0.49$  & 56446 & $18.37\pm0.20$ \\
56449 & $19.52\pm0.33$  & 56449 & $18.38\pm0.13$ \\
56456 & $19.93\pm0.28$  & 56452 & $18.75\pm0.22$ \\
\ldots & \ldots & \ldots & \ldots \\
56841 & $21.00\pm0.14$  & 56967 & $20.17\pm0.26$ \\
56846 & $20.92\pm0.11$  & 56982 & $19.86\pm0.19$ \\
56862 & $21.12\pm0.22$  & 56991 & $20.03\pm0.33$ \\
56866 & $21.32\pm0.34$  & 57007 & $19.74\pm0.12$ \\
\hline
\multicolumn{2}{c}{LRIS $g$ band} & \multicolumn{2}{c}{LRIS $r$ band}\\
MJD & Magnitude & MJD & Magnitude \\
56981 & $21.4\pm0.20$ & 56981 & $20.0\pm0.20$ \\
\hline
\end{tabular}
\caption{Selection of photometry for SN\,2009ip (the full table is available in the online version).}
\end{center}
\end{table}

\begin{figure*}
\begin{center}
\includegraphics[width=1.75\columnwidth,trim={1.2cm 0.3cm 0.5cm 0.5cm},clip]{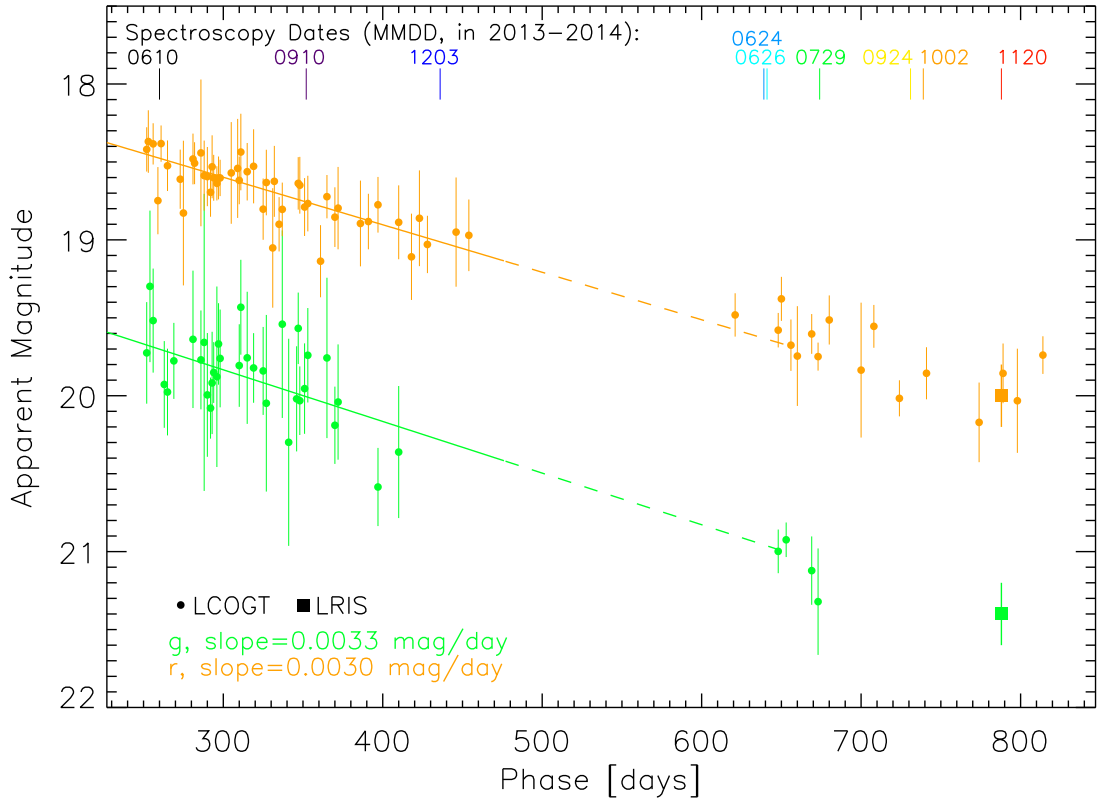}
\caption{The light curve in $g$ and $r$ to late times, in magnitudes $vs.$ the days since the start of the 2012-B event (2012-09-23). The $i$ band is not shown because SN\,2009ip was not well visible in $i$ at such late times. Solid lines are fit to the 2013 data and then extended as dashed lines to 2014. The steady linear (in magnitudes) decline seen in 2013 continued into 2014. This decline rate is slower than would be seen if $^{56}$Co were powering the light curve. Coloured vertical dashes across the top mark the dates of late-time Keck spectroscopy in 2013 and 2014 (but not the three epochs obtained in 2015). \label{fig:lc}}
\end{center}
\end{figure*}

\begin{table*}
\label{tab:spec}
\begin{tabular}{lccllcccc}
\hline
\hline
Date & MJD & Phase  & Instrument & Disperser & Slit  & Dispersion           & Range & Exposure \\
(UT) & & [days] &            &           & Width & [\AA\ $\rm pix^{-1}]$ & [\AA] & Time [s] \\
\hline
2013-06-10 & 56453 & +260 & DEIMOS & 600       & 1.0\arcsec & 0.65 & 4500-9600  &  900 \\
2013-09-10 & 56545 & +352 & DEIMOS & 600       & 1.0\arcsec & 0.65 & 4500-9600  &  900 \\
2013-12-03 & 56629 & +436 & LRIS   & 600/4000  & 1.0\arcsec & 0.63 & 3100-5600  & 1500 \\
           &  &      &        & 400/8500  & 1.0\arcsec & 1.16 & 5600-10200 & 1500 \\
2014-06-24 & 56832 & +639 & LRIS   & 600/4000  & 1.0\arcsec & 0.63 & 3100-5600  & 1400 \\
           &  &      &        & 400/8500  & 1.0\arcsec & 1.16 & 5600-10200 & 1400 \\
2014-06-26 & 56834 & +641 & DEIMOS & 1200      & 1.0\arcsec & 0.33 & 4800-7400  & 1800 \\
2014-07-29 & 56867 & +674 & LRIS   & 600/4000  & 1.0\arcsec & 0.63 & 3100-5600  & 1200 \\
           &  &      &        & 400/8500  & 1.0\arcsec & 1.16 & 5600-10200 & 1200 \\
2014-09-24 & 56924 & +731 & LRIS   & 600/4000  & 1.0\arcsec & 0.63 & 3100-5600  & 1200 \\
           &  &      &        & 400/8500  & 1.0\arcsec & 1.16 & 5600-10200 & 1200 \\
2014-10-02 & 56932 & +739 & DEIMOS & 1200      & 1.0\arcsec & 0.33 & 4800-7400  & 2700 \\
2014-11-20 & 56981 & +788 & LRIS   & 600/4000  & 1.0\arcsec & 0.63 & 3100-5600  & 1800 \\
           &  &      &        & 400/8500  & 1.0\arcsec & 1.16 & 5600-10200 & 1800 \\
2015-05-20 & 57162 & +969 & DEIMOS & 600       & 1.0\arcsec & 0.65 & 4500-9600  &  986 \\
2015-06-16 & 57189 & +996 & LRIS   & 600/4000  & 1.0\arcsec & 0.63 & 3100-5600  & 2400 \\
           &  &      &        & 1200/7500 & 1.0\arcsec & 0.40 & 5800-7400  & 2400 \\
2015-07-16 & 57219 & +1026 & LRIS   & 600/4000  & 1.5\arcsec & 0.63 & 3100-5600  & 2400 \\
           &  &      &        & 1200/7500 & 1.5\arcsec & 0.40 & 5800-7400  & 2400 \\
\hline
\end{tabular}
\caption{Late-time spectra of SN\,2009ip from Keck Observatory.}
\end{table*}

We obtained late-time photometry in the $g^{\prime}$ and $r^{\prime}$ filters with the Las Cumbres Observatory (LCO; \citealt{2013PASP..125.1031B}), starting 2014-05-26 (UT dates are used throughout this paper; YYYY-MM-DD) and continuing until 2014-12-16. Our epochs were composed of two 300--400~s exposures in each filter. We also attempted observations in $i^{\prime}$, but since the signal-to-noise ratio (SNR) is significantly lower than for $g^{\prime}$ and $r^{\prime}$, we do not include $i^{\prime}$ in this analysis. The data were reduced and calibrated to a local sequence of standard stars in Sloan Digital Sky Survey (SDSS) filters $g$ and $r$, which are very close to $g^{\prime}$ and $r^{\prime}$, as described by Graham et al. (2014). 

We also obtained one epoch of imaging with the Low Resolution Imaging Spectrometer \citep[LRIS;][]{1995PASP..107..375O, 2010SPIE.7735E..0RR} at the Keck Observatory on 2014-11-20, consisting of three 180~s exposures in the LRIS $g$ and $R$ filters. These filters have different transmission curves than SDSS $g$ and $r$, so filter conversion factors for stars and SN\,2009ip were derived. Since we use local sequence stars with $0.5 < g-r < 1.5$\,mag to determine the zeropoints of the LRIS images, we use a $T=5000$\,K blackbody spectrum to derive the filter conversion factor for stars, because this is the approximate temperature for stars in that colour range. To determine the filter conversion factor for SN\,2009ip, we use our LRIS spectrum obtained on 2014-11-20 (presented below). The conversion factors for stars and SN\,2009ip were $\sim0.05$\,mag for all filters, except for the conversion from LRIS $R$ to SDSS $r$ for SN\,2009ip, which was $-0.36$\,mag. The late-time photometry of SN\,2009ip is listed in Table \ref{tab:phot}, shown in Figure \ref{fig:lc}, and discussed further in Section \ref{sec:ana}.

We obtained a series of late-time spectra with the DEep Imaging Multi-Object Spectrograph (DEIMOS; \citealt{2003SPIE.4841.1657F}) and LRIS at the Keck Observatory from 2013-06-10 to 2015-07-16. Details of the spectroscopy, including instrument setup and exposure times, are listed in Table \ref{tab:spec}. In all cases we rotated to the parallactic angle to minimise the effects of atmospheric dispersion (\citealt{1982PASP...94..715F}; in addition, LRIS has an atmospheric dispersion corrector). All data were reduced using routines written specifically for DEIMOS and LRIS in the Carnegie {\sc Python} ({\sc CarPy}) package. The two-dimensional (2D) images were flat-fielded, corrected for distortion along the $y$ (slit) axis, wavelength calibrated with comparison-lamp spectra, and cleaned of cosmic rays before extracting the 1D spectrum of the target. This spectrum was flux calibrated using a sensitivity function derived from a standard star observed the same night in the same instrument configuration. The standard-star spectrum was also used to remove the telluric sky absorption features. 

For our analysis, we deredshift the spectra into the rest frame of the host galaxy, NGC 7259, which has a redshift of $z=0.005944$ based on $21$ cm emission\footnote{As listed in the NASA Extragalactic Database, from the \ion{H}{I} Parkes All Sky Survey Final Catalog 2006; this redshift was also used by \cite{2014ApJ...787..163G}.}. For some of our analysis in Section \ref{sec:ana} we flux calibrate spectroscopic epochs in 2015 to {\it extrapolated} photometry derived by extending the $r$-band light-curve decline rate shown in Figure \ref{fig:lc}. We chose to use the $r$ band because the flux in H$\alpha$ gives it a higher SNR, and also for consistency because the $g$ band cannot be used to calibrate the DEIMOS spectra. Our spectral time series will be publicly available in the WISEREP database \citep{2012PASP..124..668Y}\footnote{\url{http://wiserep.weizmann.ac.il}}.

\begin{figure*}
\begin{center}
\includegraphics[width=18cm,trim={0cm 1.4cm 1.5cm 2cm},clip]{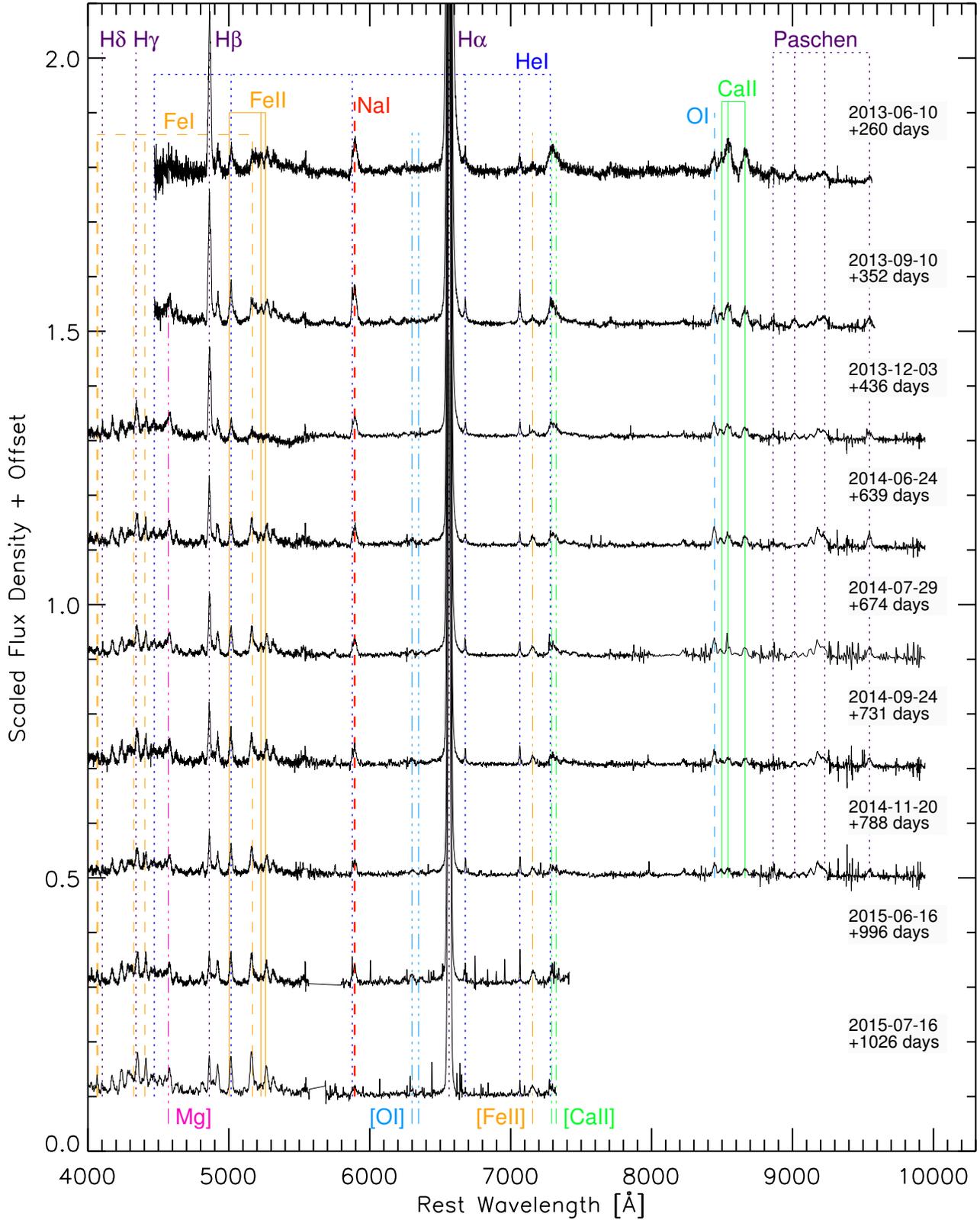}
\caption{Selected spectra of SN\,2009ip from our time series of observations, scaled and offset in flux density ($\rm erg\ s^{-1}\ cm^{-1}\ \AA^{-1}$) to aid in the comparison of the smaller features over time --- i.e., everything except H$\alpha$, which will be examined in other figures. Owing to the scaling, line strengths are not representative of true fluxes and should not be compared between epochs; instead, this plot is primarily for species identification. We have excluded three epochs: 2014-06-26 and 2014-10-02 because they were close in time to other epochs, and 2015-05-20 because it has low SNR outside of H$\alpha$. \label{fig:plot_evol_all} }
\end{center}
\end{figure*}

\begin{figure}
\begin{center}
\includegraphics[width=8.5cm,trim={0cm 0.4cm 1.5cm 1.2cm},clip]{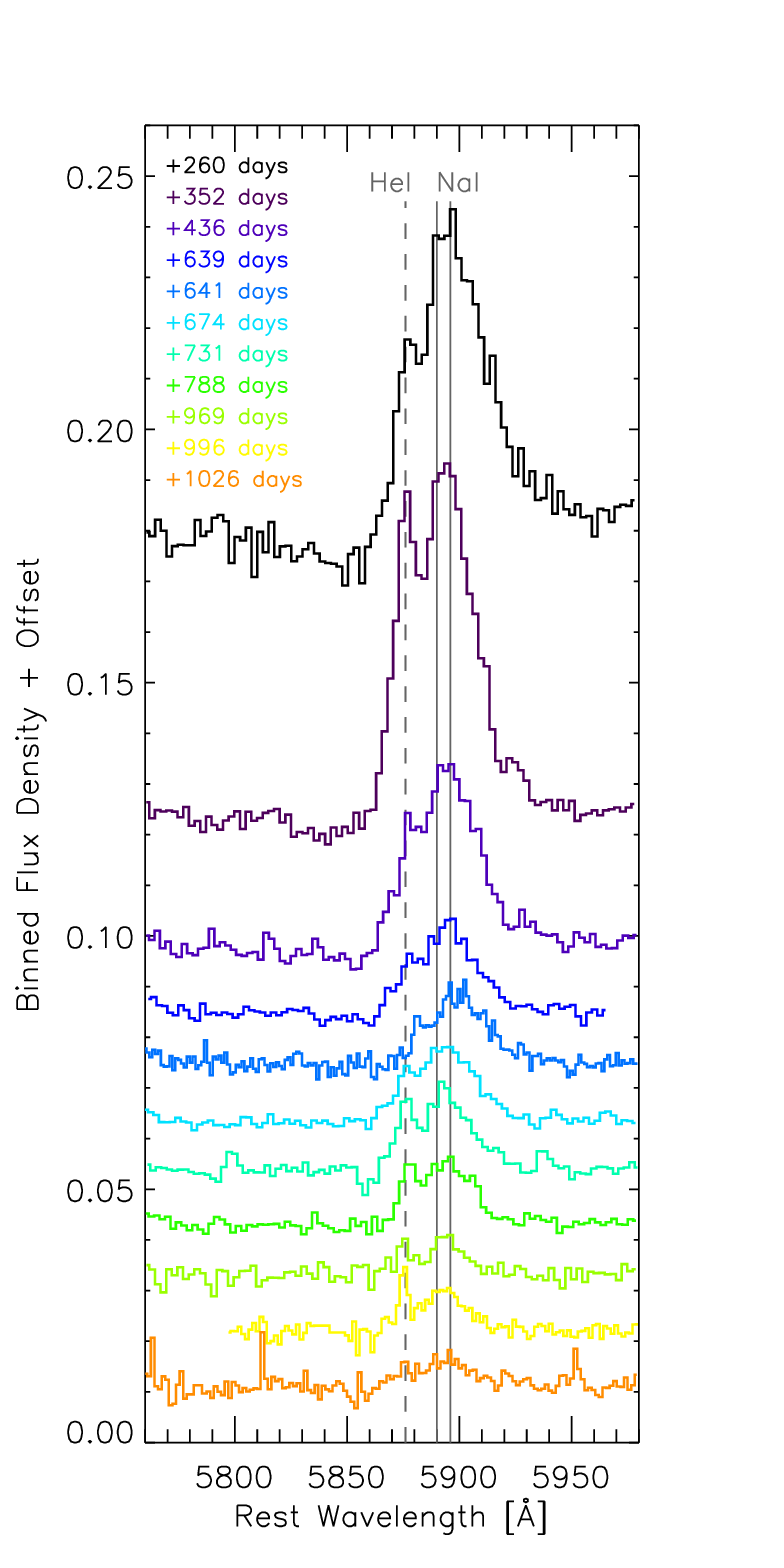}
\caption{Our spectra of SN\,2009ip in the region of the blended feature of sodium and helium at very late times. The spectra have been binned by 4 pixels in wavelength, and have arbitrary flux offsets for clarity. \label{fig:plot_evol_NaHe} }
\end{center}
\end{figure}

\section{Analysis}\label{sec:ana}

We present an analysis of our late-time photometric (\S~\ref{ssec:phot}) and spectroscopic (\S~\ref{ssec:spec}) observations of SN\,2009ip. In \S~\ref{ssec:models} we compare to the simulated light curves and spectral emission lines for models of interacting SNe presented by \cite{2016MNRAS.458.2094D}. In \S~\ref{ssec:nature} we discuss the implications of these observations regarding the true nature of SN\,2009ip.

\subsection{Photometric Decline}\label{ssec:phot}

In Figure \ref{fig:lc} we show that SN\,2009ip continued to decline throughout 2014, with a rate of $0.0033\pm0.0013$ $\rm mag\ day^{-1}$ in $g^{\prime}$ and $0.0030\pm0.0005$ $\rm mag\ day^{-1}$ in $r^{\prime}$. These values are slower than the decline rate of $^{56}$Co as seen post-plateau in core-collapse SNe like SNe\,II-P \citep{2016MNRAS.459.3939V}. This decline rate remains consistent with those reported by Graham et al. (2014), who predicted that SN\,2009ip would fade beyond its previously reported faintest magnitudes, $m_{V}=21.5$ \citep{2013ApJ...767....1P} and $m_{\rm F606W}=21.8$ \citep{2010AJ....139.1451S}, by January 2016. This late-time photometric behaviour through the end of 2014 is broadly consistent with that reported by Fraser et al. (2015), and photometry obtained in late November 2015 confirmed that SN\,2009ip had indeed declined past those faintest previous observations \citep{2015ATel.8417....1T}. However, the faintest previous observation was not necessarily a measure of the progenitor in a quiescent phase, as SN\,2009ip has been variable since 2009 \citep{2013ApJ...767....1P}, and so this photometric decline alone cannot be used as unique proof of a terminal explosion. Furthermore, as it is likely that the progenitor of SN\,2009ip was a massive binary system, it might be difficult to unambiguously associate any future variability with the same member star that produced the 2012 events. 

We extend our light curve by synthesising photometry from our Keck spectra in 2015, finding that two of the three epochs reported in Table \ref{tab:spec} produced results within several tenths of a magnitude of our extrapolation of the steady linear decline seen in Figure \ref{fig:lc}. Unfortunately, given that all three of our 2015 spectra were obtained in nonphotometric conditions owing to variable clouds, we cannot constrain our systematic uncertainties on the synthesised photometry and so are not officially including it in our analysis. Although we also obtained images with the LRIS guider camera during target acquisition, we find that they are not suitable for photometric measurements either.

The slow late-time decline rate of SN\,2009ip is similar to that observed for SNe\,IIn, such as SN\,2010jl ($0.0013\pm0.0001$ $\rm mag\ day^{-1}$ in $R$; \citealt{2012AJ....144..131Z}), SN\,2005ip ($0.0007\pm0.0002$ in $g$; \citealt{2012ApJ...756..173S}), and SN\,2006jd ($0.0037\pm0.0001$ to $0.0013\pm0.0002$ in $g$; \citealt{2012ApJ...756..173S}). We find that the late-time decline rate of SN\,2009ip is also similar to the predicted emission of SNe\,IIn at late times from the models of \cite{1994ApJ...420..268C}. They find that the total shock luminosity $1$--$2$\,yr after explosion declines at a rate of $0.0016$ $\rm mag\ day^{-1}$ (their Table 6). Not all of that energy will be emitted at optical wavelengths, so we also compare with their predicted decline rate for the luminosity in H$\alpha$, which falls in the observer-frame and rest-frame $r$ band for SN\,2009ip: $0.002$ $\rm mag\ day^{-1}$ at $1$--$2$\,yr, close to the observed $r^{\prime}$ decline rate for SN\,2009ip. Ultimately, we conclude that this continued linear decline in the photometry of SN\,2009ip at very late times is consistent with the behaviour of a SN\,IIn at similar epochs --- but it is not, on its own, a strong constraint on the true nature of the underlying explosion.

\subsection{Spectral Evolution}\label{ssec:spec}

We present our time series of Keck spectra in Figure \ref{fig:plot_evol_all}. At very late times, the spectrum remains dominated by narrow emission lines, and there is little evolution in the spectral features. This is qualitatively similar to both SNe\,IIn and SN impostors, and furthermore implies that flux contributions from a light echo or a very bright companion star are very low or nonexistent. The DEIMOS spectrum obtained on 2014-06-26 was previously published by \cite{2015MNRAS.447..772F} in their analysis of CSM-interaction SNe.

We used the atomic spectral line database produced by the National Institute of Standards and Technology (NIST; \citealt{NIST_ASD}) to identify species causing the narrow line emission in the very late-time spectra of SN\,2009ip, including hydrogen, helium, calcium, sodium, oxygen, and iron. We considered magnesium, silicon, and sulfur, but could not attribute any emission lines to these species (except for semi-forbidden \ion{Mg}{I}], discussed below). For hydrogen, we see lines in the Balmer and Paschen series within the wavelength region of our spectra. Not all Paschen lines are explicitly labeled in Figure \ref{fig:plot_evol_all}, only those that we see (the 11-, 10-, 9-, and 8-to-3 transitions). The Balmer lines are the strongest features and are considered in more detail below. We see the triplet of singly ionised calcium at $\lambda\approx8500$ \AA. Although there are other features at or near the locations of less prominent lines of neutral and singly ionised calcium (e.g., $\lambda$4227, $\lambda$7148, and $\lambda$7326 for \ion{Ca}{I}, and $\lambda$9213 for \ion{Ca}{II}), these features are better explained as being dominated by more prominent lines of other species. 

Just blueward of the \ion{Ca}{II} infrared triplet, we identify \ion{O}{I} $\lambda 8446$, which appears to strengthen over time relative to \ion{Ca}{II}. At first glance, it may seem suspicious to detect \ion{O}{I}$\lambda 8446$ without even a hint of \ion{O}{I}$\lambda 7774$; if both are produced by recombination, we expect similar relative intensities \citep{NIST_ASD}. Even if the \ion{O}{I} $\lambda 8446$ were produced by collisional excitation, we should detect weak \ion{O}{I}$\lambda 7774$. But if \ion{O}{I} $\lambda 8446$ is being produced mostly by Bowen Ly-$\beta$ fluorescence, as in the case of many Seyfert 1 nuclei and quasars (e.g., \citealt{1980ApJ...238...10G,2008ApJS..174..282L,2014AstRv...9a..29M}), then a very low intensity ratio of \ion{O}{I}$\lambda 7774$/\ion{O}{I} $\lambda 8446$ would be expected, consistent with the \ion{O}{I} $\lambda 8446$ identification. The high densities and optically thick conditions discussed in \S~\ref{sssec:balmer_emis} support this suggestion. On the other hand, an absence of oxygen features is fairly typical of SNe\,IIn while emission from CSM-interaction shrouds the SN ejecta, where the oxygen resides. Thus, it is also possible that this feature is partly  produced by \ion{Fe}{II} $\lambda\lambda8426$, 8439, 8451; a myriad of iron lines is seen, especially at the blue end of the spectrum, where we have labeled what we could identify as \ion{Fe}{I} and \ion{Fe}{II}. 

\begin{figure}
\begin{center}
\includegraphics[width=8.5cm]{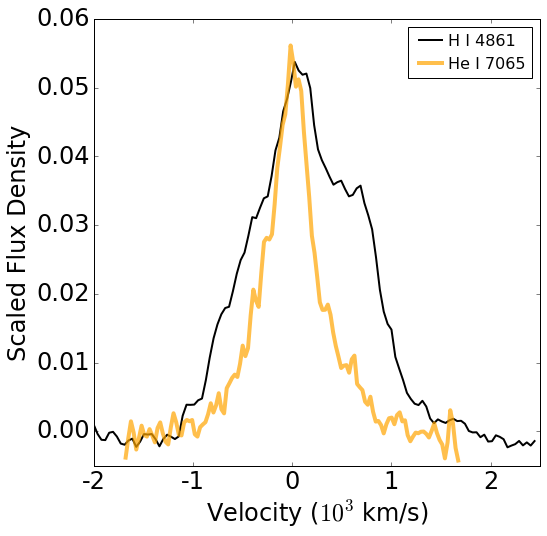}
\caption{Comparing the \ion{H}{I} $\lambda4861$ emission line (H$\beta$; thin black line) to one of the most isolated \ion{He}{I} lines, \ion{He}{I} $\lambda 7065$ (thick orange line). For this comparison we used the spectrum obtained on 2013-09-10, subtracted the pseudocontinuum determined from the flux on either side of the lines, scaled the H$\beta$ flux by a factor of $0.23$ to match the peak intensities, and converted the wavelengths into rest-frame velocity. \label{fig:H_He}}
\end{center}
\end{figure}

A closer look at the blended feature of sodium and helium is shown in Figure \ref{fig:plot_evol_NaHe}. We find that \ion{Na}{I}~D remains visible at late times, and that the P-Cygni profile seen at earlier times (e.g., \citealt{2014MNRAS.442.1166M}) is still evident, most clearly in the 2013-06-10 spectrum, but overall less pronounced than at earlier times. It also appears that the relative strength of the \ion{Na}{I}~D and \ion{He}{I} lines evolves with time, suggesting that they have different physical origins in the nebula, which is also consistent with the different line widths (clearest in the $+352$ day spectrum). A closer look at two of the more isolated emission lines of hydrogen (H$\beta$) and helium (\ion{He}{I} $\lambda7065$) is shown in Figure \ref{fig:H_He}. The H$\beta$ line is wider and appears to contain multiple blended components, whereas the \ion{He}{I} appears as a single, narrower feature. This suggests that the hydrogen emission has a more extended distribution in the nebula than He, and that the highest-velocity regions are either He poor or where \ion{He}{I} transitions are not excited.

We also see emission lines of \ion{Mg}{I}], [\ion{Ca}{II}], [\ion{Fe}{II}], and emerging [\ion{O}{I}], which are characteristically found in the nebular spectra of core-collapse SNe \citep{2014MNRAS.439.3694J}. However, since the [\ion{O}{I}] is particularly faint and appears to weaken, and because [\ion{O}{I}] is also seen in late-time spectra of CSM-interaction powered events \citep{2015MNRAS.447..772F}, we do not interpret the presence of [\ion{O}{I}] as direct evidence of a core-collapse event for SN\,2009ip. Nevertheless, we point out that weakening [\ion{O}{I}] and \ion{Na}{I}~D lines combined with \ion{Mg}{I}] lines that exhibit a relatively constant flux is consistent with a variety of progenitor mass models (see Figure 4 of \citealt{2014MNRAS.439.3694J}).

As a final note, we do not perform blackbody fits because our spectra are emission-line dominated and do not have significant SNR in the continuum. Instead, we refer the reader to the late-time bolometric analysis of optical through near-infrared (NIR) observations by \cite{2015MNRAS.453.3886F}. Since the Balmer series dominates the emission-line flux of the late-time spectra of SN\,2009ip, we now discuss the line emission and its shape in \S~\ref{sssec:balmer_emis} and \ref{sssec:balmer_asym}, respectively.

\subsubsection{Balmer-Line Emission}\label{sssec:balmer_emis}

\begin{figure}
\begin{center}
\includegraphics[width=8.5cm,trim={0.7cm 0cm 0.2cm 0.7cm},clip]{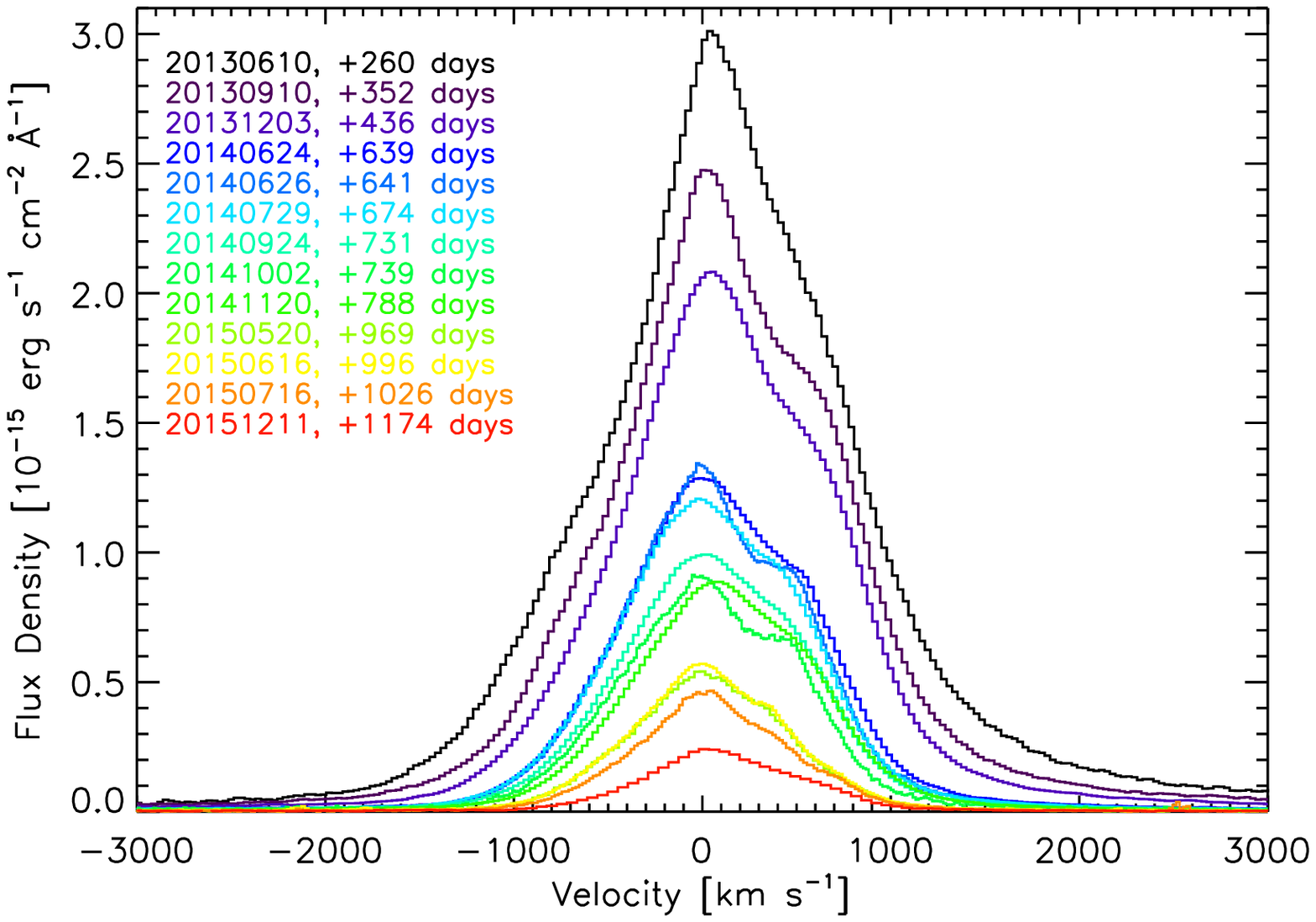}
\includegraphics[width=8.5cm,trim={0.7cm 0cm 0.2cm 0.7cm},clip]{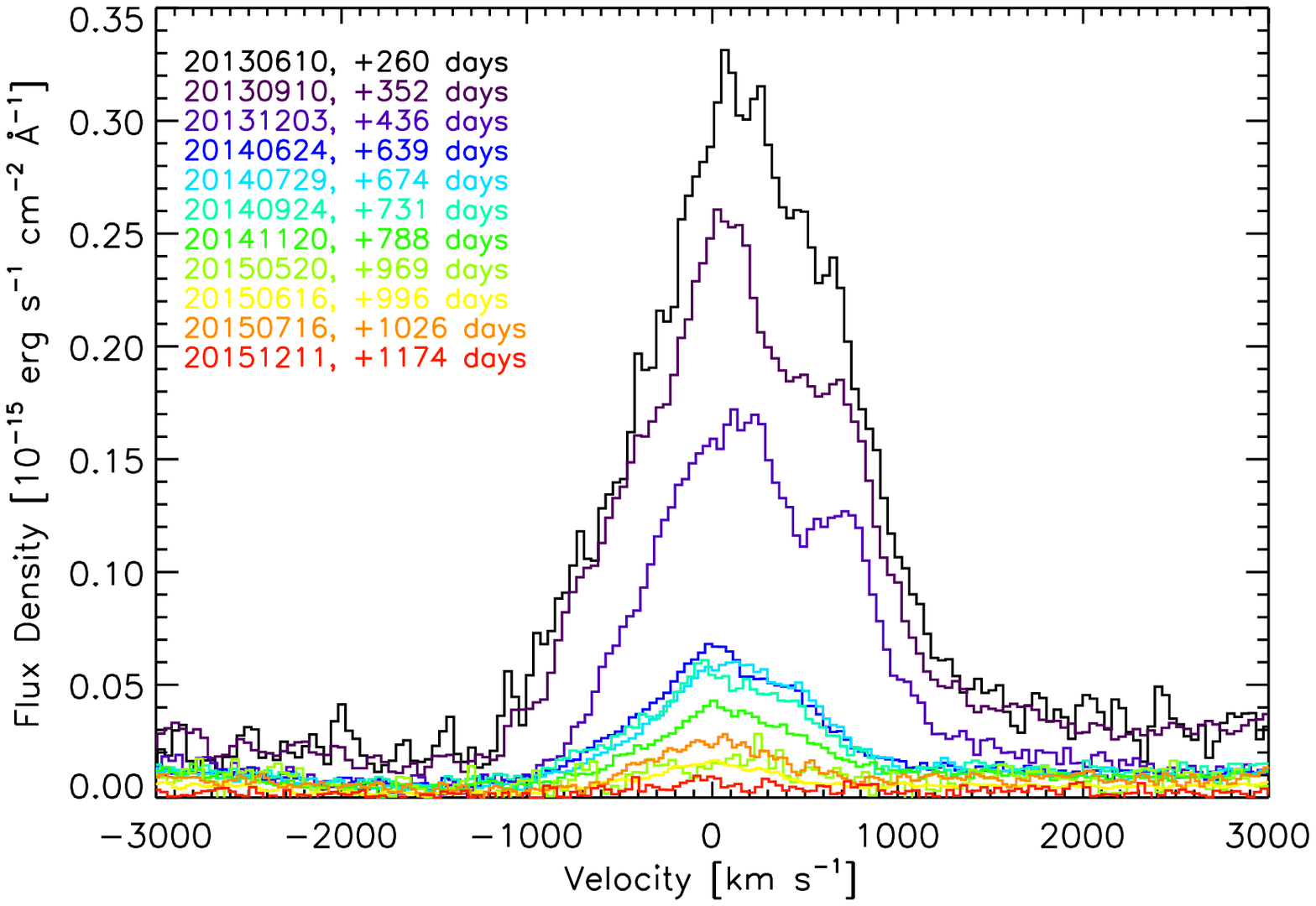}
\includegraphics[width=8.5cm,trim={0.7cm 0cm 0.2cm 0.7cm},clip]{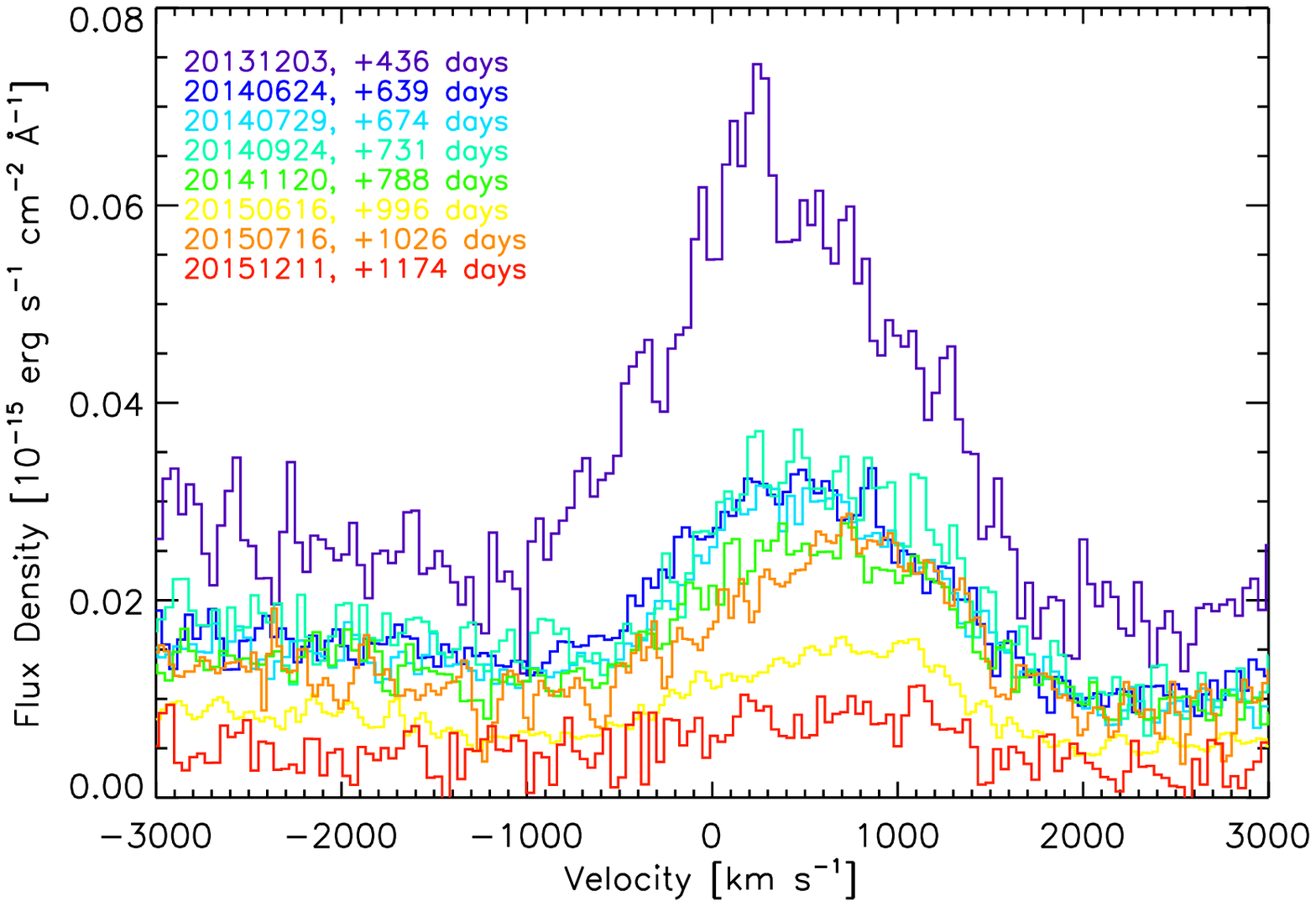}
\caption{Our series of SN\,2009ip spectra at late times for three Balmer emission lines, from top to bottom H$\alpha$, H$\beta$, and H$\gamma$. Spectra have been flux calibrated to our actual photometry or, in the case of the last three epochs, to our extrapolated $r$-band photometry. Wavelength has been converted to rest-frame velocity. \label{fig:balmer}}
\end{center}
\end{figure}

\begin{figure}
\begin{center}
\includegraphics[width=8.5cm,trim={0.7cm 0cm 0.2cm 0.7cm},clip]{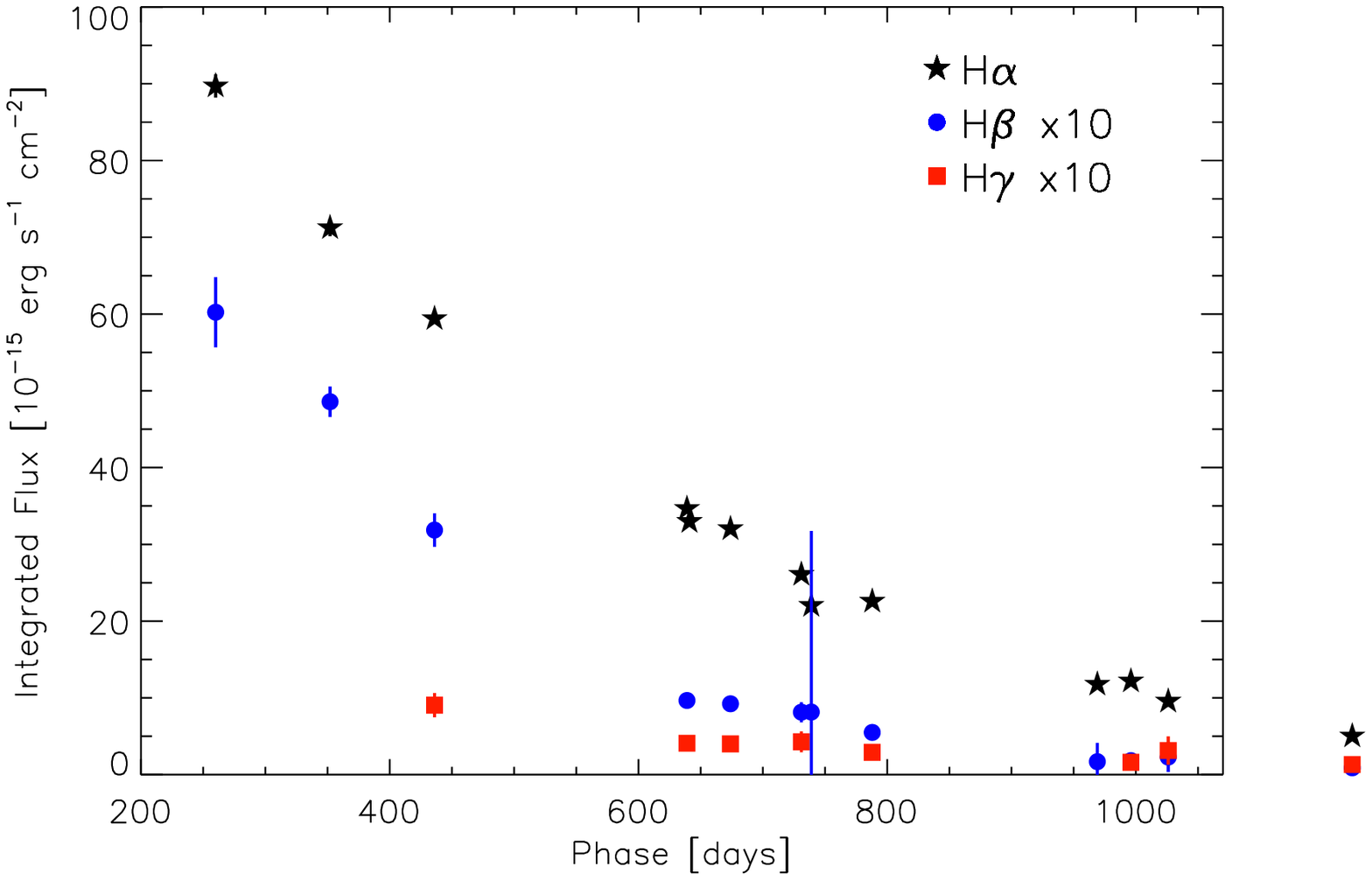}
\includegraphics[width=8.5cm,trim={0.7cm 0cm 0.2cm 0.7cm},clip]{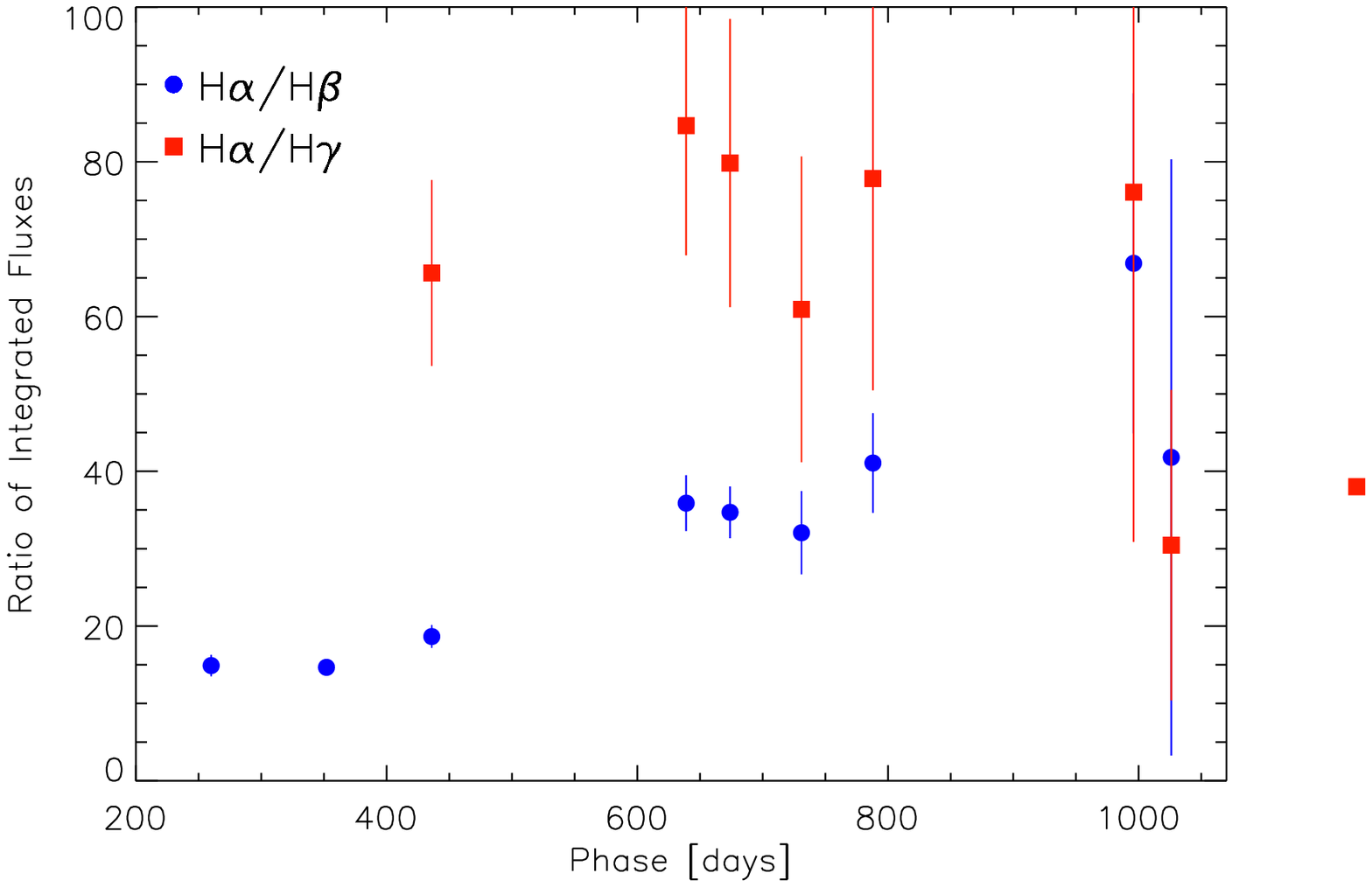}
\caption{Evolution of the continuum-subtracted integrated flux of the Balmer lines (top; note that H$\beta$ and H$\gamma$ have been scaled up to facilitate comparison), and the ratio of the integrated flux in H$\alpha$ to that in H$\beta$ and H$\gamma$ (bottom). As a reminder, the latest three points are derived from spectra that were flux calibrated to extrapolated $r$-band magnitudes, and so those integrated fluxes should be considered estimates only --- but the ratios are independent of spectral flux calibration. \label{fig:balmer_flux}}
\end{center}
\end{figure}

In Figure \ref{fig:balmer}, we zoom in on the emission lines of H$\alpha$, H$\beta$, and H$\gamma$, plotting all spectral epochs to show the evolution in luminosity and shape at late times. We find that the Balmer lines are asymmetric, exhibit multiple velocity components, and decline in flux over time. In Figure \ref{fig:balmer_flux} we plot the evolution of the integrated flux of the Balmer emission lines H$\alpha$, H$\beta$, and H$\gamma$ for SN\,2009ip, the profiles of which are shown in Figure \ref{fig:balmer}. The integrated flux is calculated by first subtracting the pseudocontinuum flux, which is determined using a linear fit to the spectral flux on either side of the emission line. We find that the integrated flux in all lines declines over time, and that the rate of decline slows down after $\sim600$ days, more so for H$\beta$ and H$\gamma$ than H$\alpha$. This is also seen in the evolution of the Balmer intensity ratio $I_{\rm H\alpha}/I_{\rm H\beta}$ (abbreviated as H$\alpha$/H$\beta$ for simplicity), which increases until $\sim600$ days and then remains nearly constant (within error bars).  

The Balmer intensity ratio H$\alpha$/H$\beta$ is sensitive to the physical state of the emitting material. \cite{2014AJ....147...23L} explored the Balmer ratio for SN\,2009ip at early times, finding that between 2012-08-30 (the peak of the precursor event, 2012-A) and 2012-10-09 (the peak of the main event, 2012-B), H$\alpha$/H$\beta$ evolved from $\sim3$ to $\sim1.5$. As they describe, a ratio of $\sim1.5$ suggests a very high density for the emitting region, $n_e > 10^{13}$ $\rm cm^{-3}$. Together with their inference of a large radiative surface area, Levesque et al. (2014) interpret the early-time H$\alpha$/H$\beta$ value of SN\,2009ip as evidence of a thin-disk geometry for the CSM.

At late-time epochs of $>600$ days, we find that H$\alpha$/H$\beta \approx 40$, which is quite high. We briefly consider whether the very late-time evolution and high value of the Balmer ratio could be due to dust formation in the nebula. Dust would cause a greater amount of extinction for blue emission lines, driving H$\alpha$/H$\beta$ to higher values. However, values of H$\alpha$/H$\beta > 10$ are not uncommon at late phases of CSM-interaction dominated events, and the ratio may have nothing to do with dust formation. For example, \cite{2008ApJ...686..467S} quote a large ratio for the luminous SN\,IIn 2006tf and suggest that line formation is no longer caused by recombination, but is excited collisionally. 

Another possibility, suggested by \cite{2015MNRAS.453.3886F}, is that the lines are experiencing ``Case C" recombination as defined by \cite{1992ApJ...386..181X} in their analysis of SN\,1987A: the Balmer continuum is optically thick, Balmer photons are immediately reabsorbed, and the higher Balmer lines ($n>4$) are resonantly trapped and decay through alternate branches such as the Paschen series. In support of this, we point out the emergence of the Paschen series at epochs $>600$ days, when the Balmer ratio is also seen to increase; moreover, as noted above, the presence of \ion{O}{I} $\lambda 8446$ produced by Bowen Ly-$\beta$ fluorescence implies high densities and large optical depths. Based on the model line emissivity in Figure 1 of \cite{1992ApJ...386..181X}, we can estimate that H$\alpha$/H$\beta \approx40$ would correspond to either $\log(\tau({\rm H}\alpha))\approx2.2$ or $4.0$. However, a value of $\log(\tau({\rm H}\alpha))\approx4.0$ predicts a relatively stronger P$\alpha$ line than was seen in the NIR spectrum shown in Figure 14 of \cite{2015MNRAS.453.3886F}, but it is unclear whether its low SNR makes it suitable for assessing a nondetection of P$\alpha$. Furthermore, a value of $\log(\tau({\rm H}\alpha))\approx2.2$ agrees with the model predictions for $\tau({\rm H}\alpha)$ at $\sim600$ days post-explosion as shown in Figure 4 of \cite{1992ApJ...386..181X}. SN\,1987A may not be completely analogous to SN\,2009ip, but we consider a similarly detailed model of the Balmer emission lines for the SN\,2009ip system beyond the scope of this work.

\begin{figure}
\begin{center}
\includegraphics[width=8.5cm]{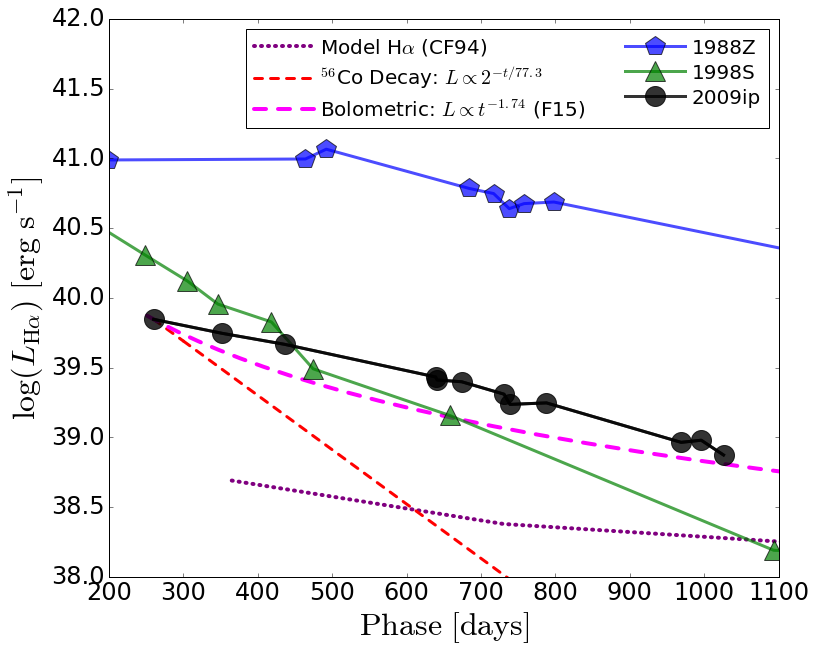}
\caption{Evolution of the intrinsic H$\alpha$ of SN\,2009ip at phases $>200$ days (black circles) compared with two SNe\,IIn: SN\,1988Z (blue pentagons; \citealt{1999MNRAS.309..343A}) and SN\,1998S (green triangles; \citealt{2001MNRAS.325..907F}). We also show the decline rate of H$\alpha$ from the models of \citet[CF94, purple dotted line]{1994ApJ...420..268C}, and the expected decline rate if $L_{\rm H\alpha}$ were driven by $^{56}$Co decay (red dashed line) or by the bolometric luminosity at late times from \citet[F15, magenta long-dashed line]{2015MNRAS.453.3886F}. The latter two have been scaled to intersect with the first late-time data point for SN\,2009ip. \label{fig:halpha_comparison}}
\end{center}
\end{figure}

Instead, we extend our evaluation of the H$\alpha$ luminosity decline in Figure \ref{fig:halpha_comparison}, where we plot the evolution in the rest-frame H$\alpha$ luminosity of SN\,2009ip at late times. For comparison, we show the late-time H$\alpha$ luminosity evolution of SN\,1988Z and SN\,1998S, which represent SNe\,IIn with faster and slower decline rates, respectively (see Figure 6 of \citealt{2017MNRAS.466.3021S}). Data for SN\,1988Z were obtained from \cite{1999MNRAS.309..343A} and converted to rest-frame luminosity using $D=70.7$ Mpc \citep{1998ApJ...500...51W}. Data for SN\,1998S was obtained from \cite{2001MNRAS.325..907F}, \cite{2004MNRAS.352..457P}, and \cite{2012MNRAS.424.2659M}, and converted to rest-frame luminosity using $D=15.5$ Mpc and a Galactic extinction of $A_V=0.68$\,mag \citep{2012MNRAS.424.2659M}. We find that the late-time evolution in $L_{\rm H\alpha}$ for SN\,2009ip is similar in decline rate to SN\,1988Z, similar in luminosity to SN\,1998S, and in general empirically similar to other SN\,IIn light curves at late times (with the caveat that SNe\,IIn are a very diverse group). 

We also show the expected decline rate if H$\alpha$ was excited by the decay of $^{56}$Co synthesized by an SN explosion, or if H$\alpha$ declines at the same rate as the bolometric luminosity at late times \citep{2015MNRAS.453.3886F}. We can see that the decline rate of H$\alpha$ is similar to the bolometric decline; the deviation is likely caused by the evolving physical state of the nebula (optical depth, density, and temperature) changing the efficiency of the emission. The late-time models presented by \citet[their Table 6]{1994ApJ...420..268C} account for the emissivity of the material and simulate the decline rate for the H$\alpha$ luminosity of a SN\,IIn at late times. We have included one of these models' results in Figure \ref{fig:halpha_comparison} and find that the model's slope is similar to the observations. The luminosity level is lower, but the luminosity is more dependent on the specific model parameters: for example, this one uses ejecta with a power-law density distribution in the outer layers, whereas ejecta with an outer-layer density distribution typical of a red supergiant star would double the H$\alpha$ emissivity \citep{1994ApJ...420..268C}. 

\subsubsection{Asymmetry of the H$\alpha$ Emission Line}\label{sssec:balmer_asym}

The degree of asymmetry in an emission line can give us clues regarding the nature of the geometry of the emitting CSM, which has been shown to be consistent with a disk or toroid via analyses of the early-time Balmer decrement \citep{2014AJ....147...23L} and time-series spectropolarimetric observations \citep{2014MNRAS.442.1166M}. In Figure \ref{fig:balmer}, the Balmer emission lines are composed of at least two velocity components, at $v \approx 0$ $\rm km\ s^{-1}$, $v \approx \pm750$ $\rm km\ s^{-1}$ (clearest in the earlier epochs of H$\alpha$, $<400$ days), and $v \approx 500$ $\rm km\ s^{-1}$ (clearest in the later epochs of H$\alpha$, $>600$ days, on the red side). The nonzero velocity components could be caused by a torus or disk of CSM released from a previous eruption, but multiple disks cannot also generate the peak at $v \approx 0$ $\rm km\ s^{-1}$ (unless one is face-on and inclined with respect to the others, but this would be a physically strange geometry for the CSM of a single system). The zero-velocity component was shown by \cite{2014MNRAS.442.1166M} to be the SN shock that appears orthogonal to the CSM geometry as it escapes the disk. If we had {\it only} our late-time data we might infer a spherical shell, which causes an emission line with a peak at $v \approx 0$ $\rm km\ s^{-1}$ and shoulders at the minimum/maximum velocity of the receding/approaching sides, similar to the red side of the H$\alpha$ line in our spectral dataset. However, a spherical shell was strongly disfavoured by early-time data (e.g., \citealt{2014AJ....147...23L}). 

\begin{figure}
\begin{center}
\includegraphics[width=8.5cm,trim={0.5cm 0.2cm 0.5cm 0.7cm},clip]{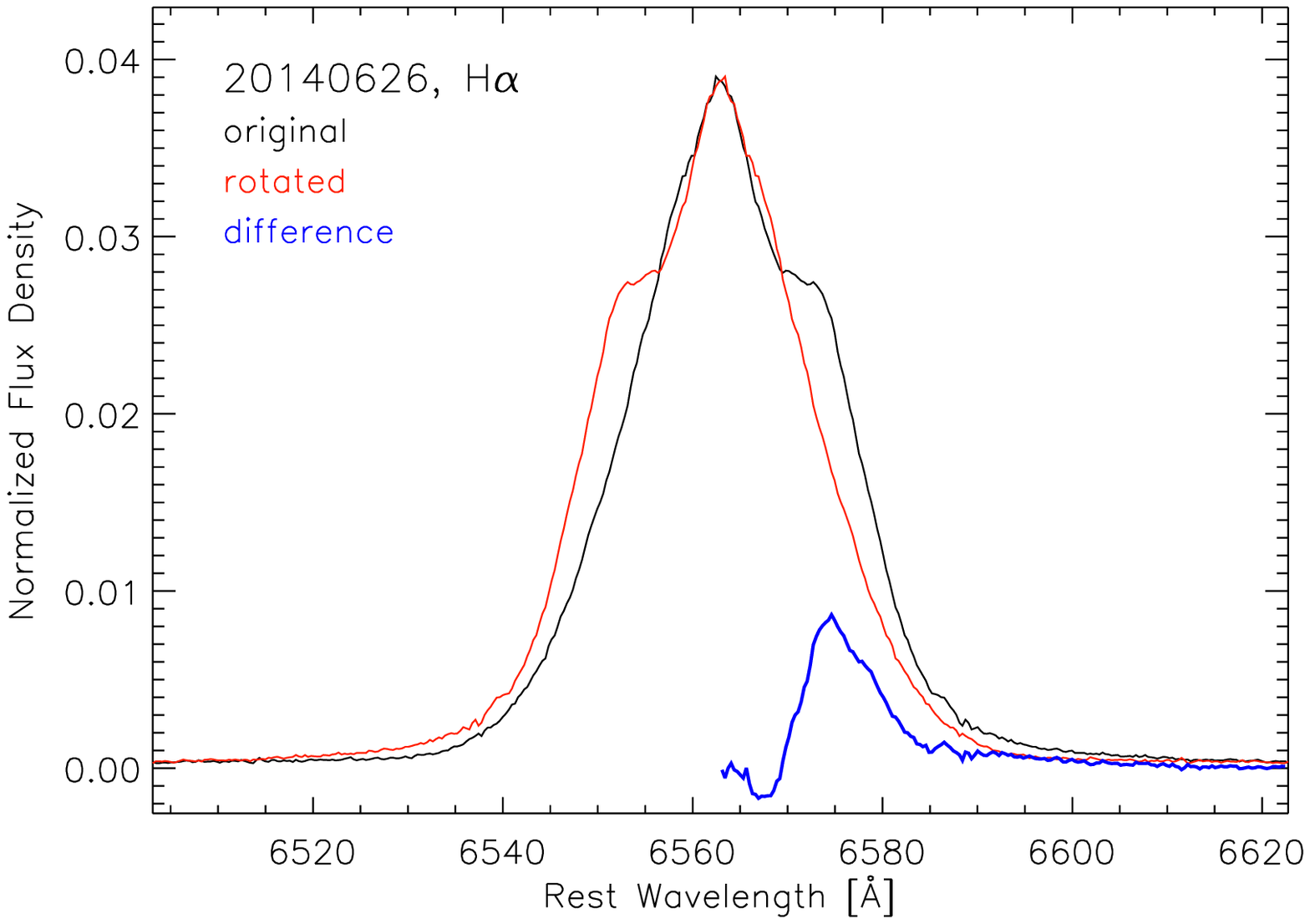}
\includegraphics[width=8.5cm,trim={0.5cm 0.2cm 0.5cm 0.7cm},clip]{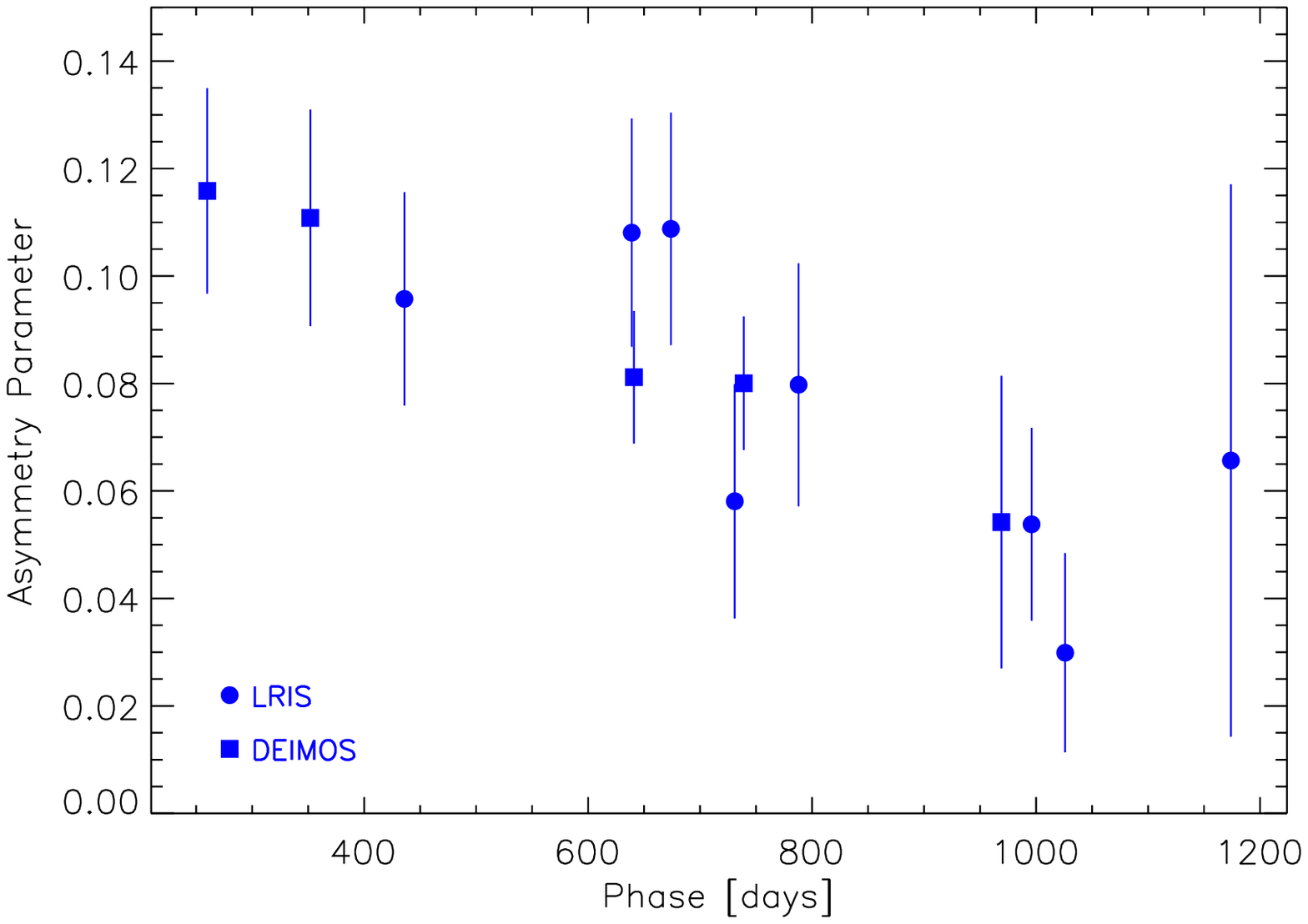}
\caption{{\it Top:} A demonstration of our derivation of asymmetry parameter $A$ using our DEIMOS spectrum from 2014-06-26, starting with the original rest-frame spectrum in the region of H$\alpha$ ($f(\lambda)$, black), rotating it around the peak wavelength ($f_{\rm rot}(\lambda)$, red), and calculating the absolute value of the difference ($\Delta f = f-f_{\rm rot}$, blue). The asymmetry parameter $A$ is the integral over wavelength of $\Delta f$, or the total area under the red line. {\it Bottom:} The asymmetry parameter $A$ as a function of light-curve phase for SN\,2009ip at late-time epochs. The amount of asymmetry in the H$\alpha$ line appears to decrease with time, and always shows a red-side excess (i.e., $A>0$). \label{fig:balmer_asym}}
\end{center}
\end{figure}

The amount of asymmetry in the H$\alpha$ emission line as a function of time is difficult to gauge by examining Figure \ref{fig:balmer}, and so we have come up with a parameterisation to quantify the amount of asymmetry. For all epochs of our rest-frame spectra, we find the peak of the H$\alpha$ line (the method to identify the peak is described below), extract the region within $\pm60$ \AA, and normalise the spectral segment to have an integrated flux equal to unity. We then rotate the spectrum around the peak and subtract the rotated from the original to create a difference spectrum, $\Delta f = f - f_{\rm rot}$. This is demonstrated in Figure \ref{fig:balmer_asym}. The integral of the difference spectrum over the red half of the line is our asymmetry parameter,

\begin{equation}
A = \int_{\lambda_{\rm peak}}^{\lambda_2}{\Delta f \ d\lambda},
\end{equation}

\noindent where $\lambda_2 \approx \lambda_{\rm peak} + 60$ \AA. Since we first normalise to a total integrated flux of unity, the absolute value of the $A$ parameter can be thought of as a fraction of the line flux that is asymmetric, and $A$ parameters from different epochs can be compared. Values of $A>0$ indicate that the red side of the line has an excess of flux.

To compensate for the fact that some spectra are noisier than others, we do not simply use the pixel with the peak flux value, but adopt a more complicated process that is based on an assumption of symmetry in the very centre of the H$\alpha$ peak. First, we smooth the spectrum with a Savitzky-Golay filter of width $5$ pixels (i.e., the {\sc idl} astronomy library function {\sc poly\_smooth}). We then extract a small spectral region of the $5$ pixels at peak flux, and calculate $\sum |f-f_{\rm rot}|$ for this tiny segment. We repeat this four times using the $\pm2$ adjacent pixels as the peak flux pixel, and use the pixel which minimizes $\sum |f-f_{\rm rot}|$ as the true peak. We find that the identified wavelength of peak flux varies randomly within a standard deviation of $0.75$ \AA\ (1--2 pixels, with an average consistent with $v\approx0$ $\rm km\ s^{-1}$), and does not show any monotonic variation with time. The asymmetry parameter $A$ is calculated using this identified peak pixel and the original, unsmoothed flux spectrum. The uncertainty in $A$ is evaluated to be the difference between this value of $A$ and the value returned if we substitute the next-best peak pixel (which is always adjacent). In other words, our uncertainty in $A$ is dominated by the uncertainty in the wavelength of the line peak, not from the flux SNR.

In Figure \ref{fig:balmer_asym}, we plot the evolution of the H$\alpha$ line's $A$ parameter as a function of time for our spectra. The asymmetry fluctuates over time, showing a generally decreasing trend (i.e., becoming more symmetric) as the flux declines. This does not necessarily mean the geometry of the CSM is changing, but that the location in the CSM producing most of the H$\alpha$ photons may be moving from a more to less asymmetric configuration. One plausible scenario is that the most asymmetric material has the highest density and cools fastest, but in this complex environment the CSM emission may also be influenced by a reverse shock or an internal power source (e.g., nuclear decay products in the case of a genuine SN, or the post-eruption stellar product in the case of a nonterminal event). We consider a detailed model of the CSM and its post-impact cooling beyond the scope of this paper (instead, see \citealt{2015MNRAS.453.3886F} for this type of modeling and analysis).

Finally, we note that the asymmetry in the H$\alpha$ line does not exhibit either of the typical behaviours associated with dust formation, which is an increasing amount of absorption on the red wing or an increasingly blueshifted peak wavelength owing to far-side photons being extinguished by a higher column density of dust on their path toward the observer. \cite{2015MNRAS.453.3886F} also remark on the lack of this typical signature of dust in their late-time spectra, and suggest that substantial dust formation is not occurring in the SN\,2009ip system. However, these are the effects when the dust is being formed within the region that also contains the material creating the emission line --- in other words, when the absorbing dust and emitting gas are colocated. SN\,2009ip has multiple velocity components in its Balmer lines and the situation could be more complicated. For example, if the emitting gas is located entirely within the dust-formation region, there would not be a large difference in the dust column density along the line of sight through the emission-line region. However, as final piece of evidence against dust formation, we note that our observed photometric decline is not significantly more rapid in the $g$ band than in the $r$ band as would be expected with dust formation. Ultimately, we conclude that we see no signature of dust formation in the system of SN\,2009ip.

\subsection{Qualitative Comparison to Models}\label{ssec:models}

\cite{2016MNRAS.458.2094D} present simulated light curves and spectra for events with CSM interaction for a variety of parameters describing the energy, mass, and distribution of the ejecta and CSM. These simulations are mainly for phases of $<200$ days and are aimed to reproduce the observations of other SNe\,IIn such as SN\,1994W and SN\,1998S, which limits their applicability to the late-time dataset we focus on in this paper; however, we find that a broad, qualitative comparison of the observations at phases $<300$ days presented by \cite{2014ApJ...787..163G} and in this work to these models can provide some clues regarding the nature of SN\,2009ip.  

We find that the simulated bolometric light curves for models ``A," ``B3," ``R1," and ``R3" of \citet[their Figures 2 and 8]{2016MNRAS.458.2094D} are similar to the bolometric light curve at $<200$ days for SN\,2009ip presented in Figure 11 of \cite{2014ApJ...787..163G}. They are similar both in terms of peak bolometric luminosity ($\sim10^{43}$ $\rm erg\ s^{-1}$) and light-curve features: a quick rise to peak within 20 days followed by smooth drops before settling on a shallower late-time decline rate $>150$ days. Alternative models ``C" and ``D," which \cite{2016MNRAS.458.2094D} developed to match the light-curve observations for SN\,IIn 1994W, do not resemble SN\,2009ip as well: they exhibit a delayed rise to peak, followed by a plateau or slow decline and then a steep drop in luminosity around $200$ days (their Figures 11 and 18; unfortunately, the bolometric evolution after 200 days is not shown, so it is unclear whether the light curve settles onto a shallow decline). The main physical differences are that their models ``C" and ``D" contain a less massive ($0.3$ $\rm M_{\odot}$) and less energetic ($<0.08\times10^{51}$ $\rm erg$) ejecta component interacting with a very massive ($6$ $\rm M_{\odot}$) outer shell that was previously ejected in a single eruption event of the star, while models ``A,", ``B3," ``R1," and ``R3" contain a more massive (1--10 M$_{\odot}$) ejecta component with a kinetic energy up to that of a typical supernova ($0.05$--$1\times10^{51}$ $\rm erg$) interacting with a less massive ($0.1$--$1$ $\rm M_{\odot}$) shell of wind-driven CSM. This points to SN\,2009ip being more similar to a regular SN\,IIn than to a low-mass ejection event.

\cite{2014ApJ...787..163G} documented the evolution in H$\alpha$ for SN\,2009ip, showing how the narrow component (full width at half-maximum intensity [FWHM] $\approx 1000$ $\rm km\ s^{-1}$) rose and fell with the bolometric luminosity, and the broad component (FWHM $\gtrsim 5000$ $\rm km\ s^{-1}$) appeared after peak brightness and then disappeared around the time the light curve settled onto a shallower decline rate (e.g., their Figure 17). As we have shown here, this shallow decline rate and the narrow component of H$\alpha$ persist to late times, $>200$ days. Among the models of \cite{2016MNRAS.458.2094D} that we identified as having bolometric light curves matching that of SN\,2009ip (``A," ``B3," ``R1," and ``R3"), model ``R1" exhibits narrow H$\alpha$ emission at all epochs, like SN\,1994W, which {\it never} exhibited a broad component of H$\alpha$ in observations up to $\sim120$ days \citep{1998ApJ...493..933S}. This leaves models ``A," ``B3," and ``R3" as matching both the light curve and spectra, and these are the three models with ejecta energies similar to that of a typical SN, $10^{51}$ $\rm erg$.

We can also compare to the models by \cite{2015MNRAS.449.4304D}, who synthesize light curves and spectra for superluminous SNe\,IIn with a variety of physical CSM parameters: in their Figure 20, they show the simulated bolometric light curves for their models out to $+450$ days. Past $200$ days, most of the light curves experience a sudden and significant drop in luminosity that we did not observe for SN\,2009ip, except for three ({\sc Xe3m6r, Xm3}, and {\sc Xm6}). These models are distinguished from the others by having progenitor stars that experienced a more rapid mass-loss rate during the late stages of stellar evolution, and in one case, a CSM that extends to a radius $50\%$ farther than the others. The total bolometric luminosities for these models are $\gtrsim3$ orders of magnitude higher at late times than SN\,2009ip ($\log{L_{\rm bol}}\approx40$; \citealt{2015MNRAS.453.3886F}) owing to a more energetic underlying explosion driving the interaction. The synthesized early-time spectra for these three models are also qualitatively similar to those of SN\,2009ip (see Figure 22 of \citealt{2015MNRAS.449.4304D}). Late-time synthesized spectra are not shown for these three models, but are given for their model X, in a comparison to luminous SN\,IIn 2010jl (see also Section \ref{ssec:nature}). The synthetic spectra exhibit significantly more asymmetry and have a strong, blueshifted component of the H$\alpha$ line, which is even more exaggerated than that seen for SN\,2010jl at $\sim200$ days. \cite{2015MNRAS.449.4304D} explain this blueshifted feature as arising from an optically thick, cool dense shell (CDS) between the reverse and forward shocks in the CSM, but for SN\,2009ip we would not expect to see this signature because we found that the CSM exterior to the shock front is not yet optically thin (i.e., Section \ref{sssec:balmer_emis}). Although not every aspect of the models by \cite{2015MNRAS.449.4304D} represent the physical scenario of SN\,2009ip, where appropriate comparisons can be made they support a moderately energetic explosion into a large and extended mass of CSM. 

There is, however, an additional perspective to consider for SN\,2009ip: that the broad component of H$\alpha$ is from the SN ejecta and entirely unassociated with the CSM. The scenario is that the SN light can escape and contribute to the spectra at all epochs owing to the inclined ring structure of the CSM \citep{2014AJ....147...23L,2014MNRAS.442.1166M}, but is swamped by the CSM-interaction emission during the peak of the 2012-B event and at later times, making it visible only during the decline of the 2012-B event and disappearing at a phase of $\sim100$ days \citep{2014MNRAS.438.1191S,2014ApJ...787..163G}. In this scenario, the H$\alpha$ emission from the CSM interaction could have been narrow at all epochs, similar to SN\,1994W, and the ejecta need not have the energy of a typical SN (model ``R1'' has $E_{\rm kin,ejecta} = 5\times10^{49}$ $\rm erg$) and/or be rapidly decelerated by a massive CSM \citep{2016MNRAS.458.2094D}. This scenario of an inclined perspective on an asymmetric, multi-zone system also helps to reconcile our conflicting late-time observations of an optically thick CSM combined with emission lines from an SN nebula such as oxygen and iron, with helium having a remarkably different line profile (Figure \ref{fig:H_He}). Ultimately, without detailed modeling of the spectral evolution resulting from interaction with an aspherical distribution of CSM, as has been inferred for SN\,2009ip, it remains difficult to draw any definitive conclusions on the energetics and nature of the explosion powering the events from 2012.

\begin{figure}
\begin{center}
\includegraphics[width=8.5cm]{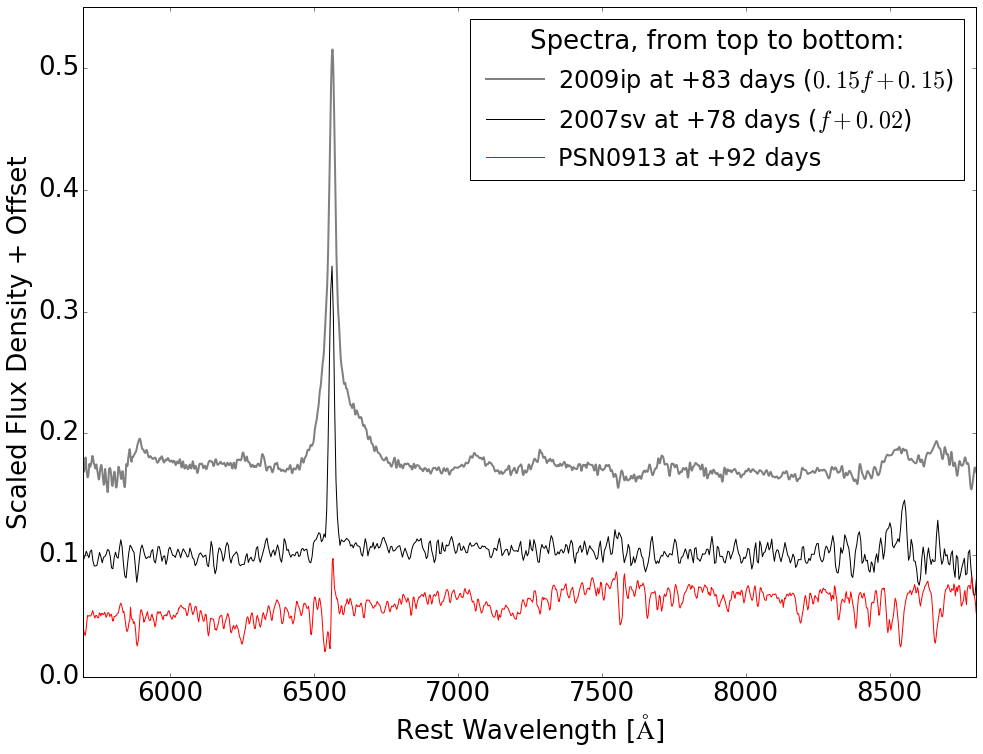}
\caption{A spectrum of SN\,2009ip at $+82$ days (grey, \citealt{2014ApJ...787..163G}) compared to two SN impostor events at similar phases: SN\,2007sv (black, \citealt{2015MNRAS.447..117T}) and PSN\,J09132750+7627410 (red, \citealt{2016ApJ...823L..23T}). Spectra have been converted to rest-frame redshifts and had their flux scaled and/or offset for clarity. \label{fig:compimp} }
\end{center}
\end{figure}

\begin{figure}
\begin{center}
\includegraphics[width=8.5cm]{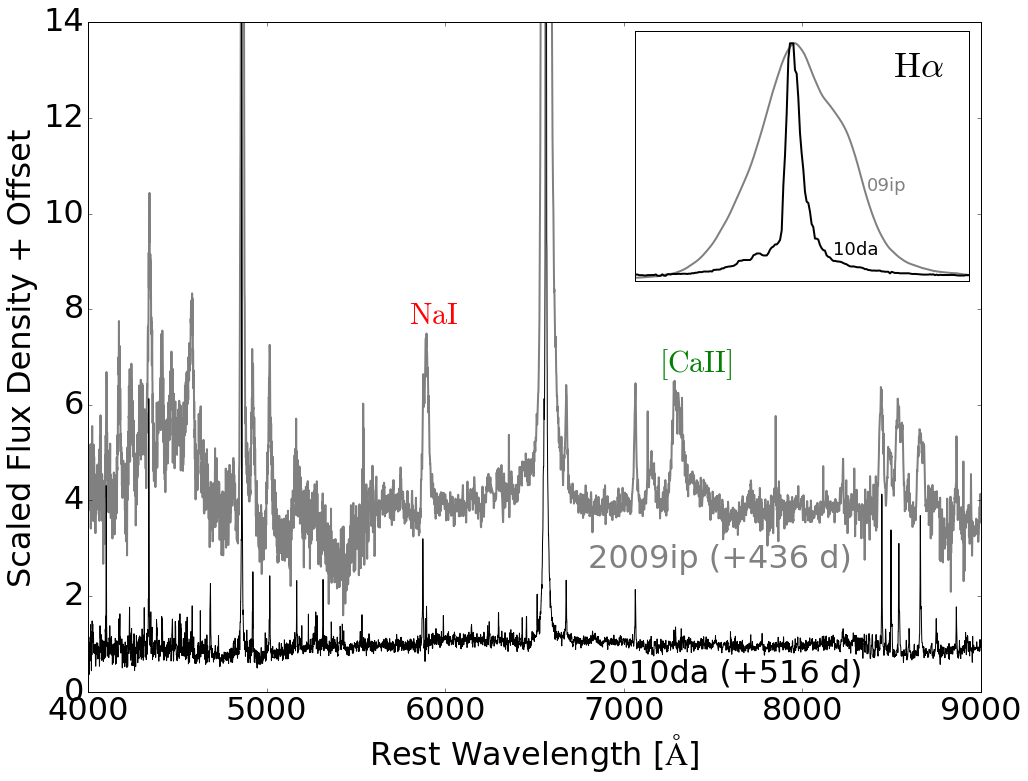}
\caption{Our spectrum of SN\,2009ip at $+436$ days (grey) compared with SN impostor SN\,2010da at $+516$ days after discovery (black; \citealt{2016ApJ...830...11V}). All emission lines in the impostor spectrum are narrower than those of SN\,2009ip, and the inset shows this in detail for H$\alpha$. Additionally, the spectrum of SN\,2010da does not exhibit the blue-end forest of iron lines, nor \ion{Na}{I} or [\ion{Ca}{II}]. \label{fig:2010da} }
\end{center}
\end{figure}

\begin{figure}
\begin{center}
\includegraphics[width=8.5cm,trim={0.6cm 0.2cm 1.7cm 1.5cm},clip]{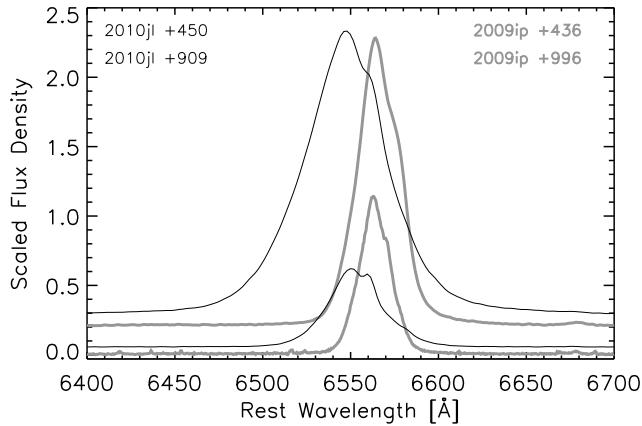}
\caption{Comparing the H$\alpha$ emission in our late-time spectra of SN\,2009ip (thick grey lines) to SN\,IIn 2010jl spectra at similar phases (thin black lines, from \citealt{2016MNRAS.456.2622J}). The flux densities ($\rm erg\ s^{-1}\ cm^{-1}\ \AA^{-1}$) have been scaled to facilitate the comparison of features. The major difference between SN\,200ip and SN\,IIn 2010jl is the wider Balmer emission in SN\,2010jl. \label{fig:jencsonHa} }
\end{center}
\end{figure}

\begin{figure*}
\begin{center}
\includegraphics[width=17cm,trim={0.9cm 0.2cm 2.5cm 0.8cm},clip]{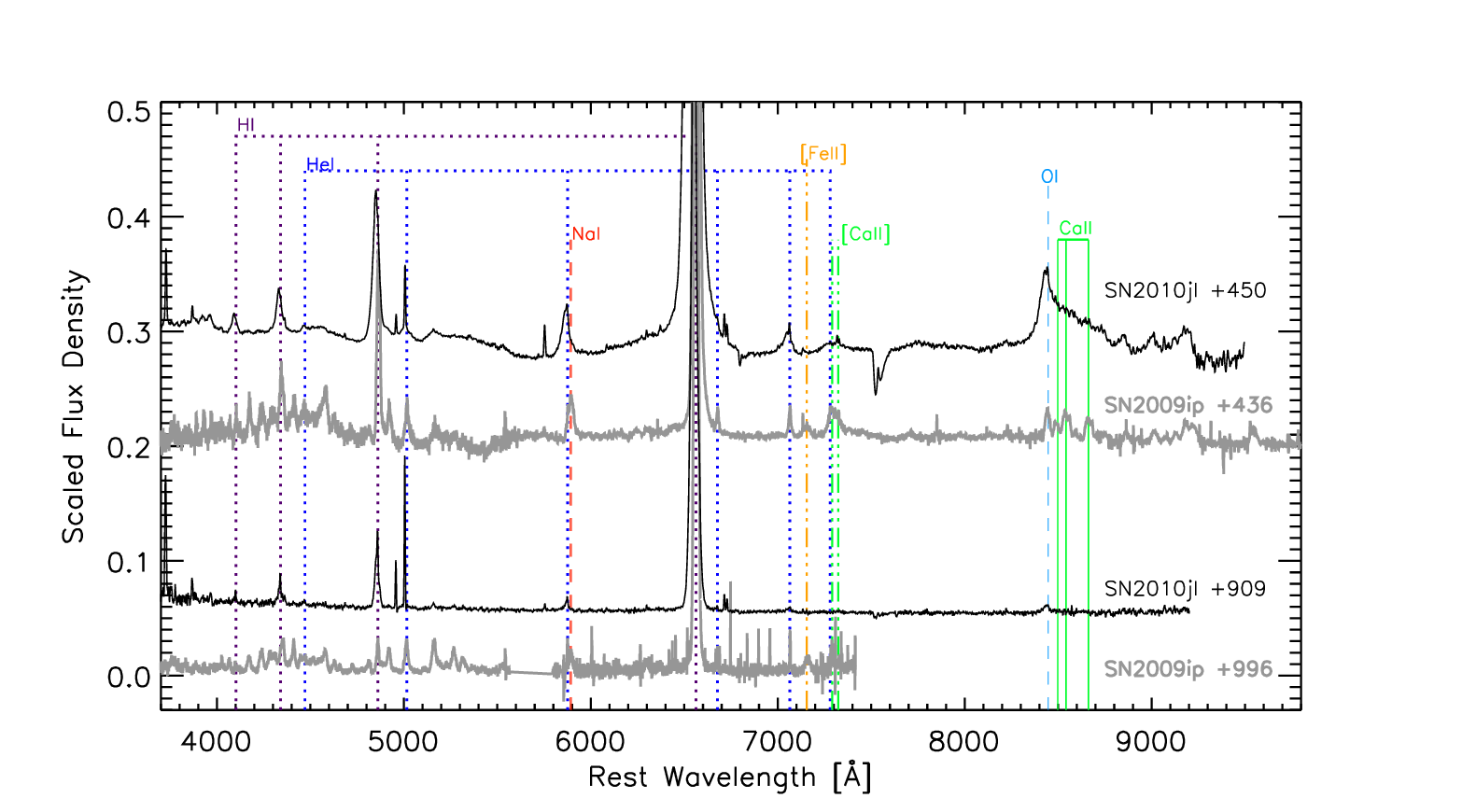}
\caption{A comparison of two of our late-time spectra of SN\,2009ip (thick grey lines) to SN\,IIn 2010jl spectra at similar phases (thin black lines, from \citealt{2016MNRAS.456.2622J}). The flux densities ($\rm erg\ s^{-1}\ cm^{-1}\ \AA^{-1}$) have been scaled to facilitate the comparison of features.  We show the full spectrum with some emission-line species labeled; the major difference between SN\,2009ip and SN\,IIn 2010jl is the stronger \ion{O}{I} line in SN\,2010jl. \label{fig:jencson} }
\end{center}
\end{figure*}

\subsection{True Supernova or Impostor?}\label{ssec:nature}

In Section \ref{sec:intro} we described recent attempts to characterize the true nature of 2009ip with late-time observations such as those of \cite{2014MNRAS.438.1191S}, who compared the late-time spectra of SN\,2009ip and SN\,2010mc to that of SN\,1987A, interpreting all three as the core collapse of blue supergiant stars with different distributions of CSM. \cite{2014ApJ...787..163G} obtained a late-time spectrum of SN\,2009ip with the Gemini Observatory on 2013-06-12, which is a phase of $+260$ days after the start of the 2012-B event. This is just two days later than the earliest Keck spectrum presented in this work, which focuses on the late-time evolution. In their Figure 19, \cite{2014ApJ...787..163G} compare their late-time Gemini spectrum to two spectra of SN\,2009ip during its LBV phase in 2009 to demonstrate that it was not returning to its earlier quiescent state as it might if it was a nonterminal outburst, but remained significantly different. They also compared their $+260$ day spectrum to representatives of typical Type IIn and II-P (plateau) SNe and to the SN impostor SN\,2008S. They find that the late-phase spectrum of SN\,2009ip does not resemble a typical SN\,II-P, which exhibits broad features and more prominent oxygen and calcium lines (associated with the SN ejecta). 

\cite{2014ApJ...787..163G} show that their late-phase spectrum of SN\,2009ip does resemble that of SN impostor SN\,2008S, but that SN\,2009ip exhibits broader emission lines for all species and that there is more flux in the \ion{He}{I} lines of SN\,2009ip. They also illustrate that the heterogeneity of features in late-time SN\,IIn spectra almost prohibits drawing any conclusions. For example, the $\sim300$ day spectrum of peculiar SN\,IIn 1998S exhibits a very broad H$\alpha$ line with three distinct velocity components that are not seen in the $+262$ day spectrum of SN\,2009ip, and while the $+115$ day spectrum of SN\,IIn 1988Z has emission lines of a similar shape, its \ion{He}{I}, \ion{Fe}{III}, and \ion{Ca}{II} are all weaker than in SN\,2009ip. Here we extend this spectral comparison to SNe\,IIn and impostors, neglecting SNe\,IIP since \cite{2014ApJ...787..163G} already showed they were a poor match, and choosing recently published data for our comparison objects. We discuss three categories in turn: SN impostors, events that resemble SN\,2009ip, and SNe\,IIn. 

\textbf{SN Impostors --} It is difficult to find spectroscopic data for SN impostors at phases $>200$ days. Instead, in Figure \ref{fig:compimp} we compare spectra of two impostor events, SN\,2007sv and PSN\,J09132750+7627410 at $\sim80$ days\footnote{Obtained from WISEREP, \url{http://wiserep.weizmann.ac.il}.} with the $+82$ day spectrum of SN\,2009ip presented by \cite{2014ApJ...787..163G}. The presence of broad H$\alpha$ in SN\,2009ip, but not in SN\,2007sv nor PSN\,J09132750+7627410, is clearly seen, and is the main difference between SN impostors and true SNe. A lack of (or very narrow) emission lines of He and Ca is another distinguishing characteristic of impostor events that is demonstrated by Figure \ref{fig:compimp}. 

The SN impostor SN\,2010da is one of the few to have been spectroscopically monitored to a very late phase of 1881 days past discovery, as presented by \cite{2016ApJ...830...11V}. The late-time spectra of SN\,2010da are qualitatively similar to ours of SN\,2009ip, exhibiting minimal evolution in features and with a similarly large Balmer ratio at late times, and three velocity components similar to H$\alpha$ (e.g., see Figure 11 from \citealt{2016ApJ...830...11V}). However, the spectra have several important differences. In Figure \ref{fig:2010da} we compare our spectrum of SN\,2009ip with a late-phase spectrum of SN\,2010da at $\sim500$ days (with respect to the onset of the 2012-B event for SN\,2009ip, and with respect to the discovery epoch for SN\,2010da). We can see that although some similar species are present (mainly H, He, and Ca), all of the emission lines in SN\,2010da are significantly narrower than those in SN\,2009ip, and that although Fe, Na, and [Ca~II] are all significant in the SN\,2009ip spectrum, they are not at all present in SN\,2010da. Furthermore, SN\,2010da differed from SN\,2009ip by reaching a fainter peak brightness and exhibiting a steady increase in optical brightness from 200 to 2000 days past the initial peak, clearly indicating that its main event was nonterminal. In addition, \cite{2016MNRAS.457.1636B} show that SN\,2010da became a variable X-ray source at late times; they suggest that the system is composed of either a LBV plus neutron-star binary or a single massive star, and did not experience a terminal SN explosion. 

\textbf{SN\,2009ip-like Events --} \cite{2014MNRAS.438.1191S} first noted the similarity between SN\,2009ip and SN\,2010mc, which also exhibited a precursor event, and suggested they were both genuine SNe from comparable progenitor systems. Since then, more objects have been added to this family of SNe\,IIn. A year of photometric and spectroscopic observations of the SN\,2009ip-like event SN\,2015bh was presented by \cite{2016MNRAS.463.3894E} and \cite{2016arXiv160900731G}. At early times, the light curve of SN\,2015bh exhibited a precursor event followed by a second rise to a brighter peak, and at late times it exhibited a steady decline of $0.0038$ $\rm mag\ day^{-1}$ and $0.0046$ $\rm mag\ day^{-1}$ in the $g$ and $r$ bands at phases of $>150$ days \citet[their Table 1]{2016MNRAS.463.3894E} --- similar to SN\,2009ip in both aspects. The spectroscopic evolution of SN\,2015bh is also similar to that of SN\,2009ip, and at late times SN\,2015bh exhibits emission lines of H, He, Na, Ca, O, and Fe, as well as a large $L_{\rm H\alpha}/L_{\rm H\beta}$ ratio, like SN\,2009ip (the latest spectrum is presented by \citealt{2016arXiv160900731G} at $+374$ days). At late times the asymmetry in the H$\alpha$ emission line becomes more prominent; the multiple velocity components are more distinct than are seen for SN\,2009ip, but similarly attributable to an asymmetric geometry of the CSM \citep{2016MNRAS.463.3894E, 2016arXiv160900731G}. LSQ13zm is another recent example of an object that is similar to both SN\,2009ip and SN\,2015bh \citep{2016MNRAS.459.1039T}.

\textbf{Type IIn SNe --} In this final category, we compare our observations of SN\,2009ip to the recently published late-time spectral evolution of SN\,IIn\,2010jl \citep{2016MNRAS.456.2622J}. In Figures \ref{fig:jencsonHa} and \ref{fig:jencson}, we compare two of our late-time SN\,2009ip spectra, at phases of +450 and +909 days, to spectra of SN\,2010jl at similar phases of +436 and +996 days from \cite{2016MNRAS.456.2622J}\footnote{Spectra were downloaded from WISEREP, \url{http://wiserep.weizmann.ac.il}}. The main differences between the SN\,IIn spectrum and SN\,2009ip are that the SN\,IIn has stronger \ion{O}{I} and broader Balmer emission lines at late times than SN\,2009ip. The former could be attributed to intrinsic differences in the CSM's ionisation field, density, and/or metallicity; the latter suggest either intrinsically higher velocity CSM or a larger amount of kinetic energy imparted by the ejecta. It is notable that the H$\alpha$ emission of SN\,IIn\,2010jl also exhibits multiple velocity components --- although a spherically symmetric CSM could generate this feature, polarimetry of SN\,2010jl has shown that it, too, had an asymmetric explosion and CSM environment \citep{2011A&A...527L...6P}. 

In summary, we find that the late-time spectral evolution of SN\,2009ip, being dominated by the lingering emission of impacted CSM, is generally more similar to that of SNe\,IIn than to SN impostors. However, given that both of these are heterogeneous groups and that we found only one SN impostor with which to compare at very late phases, we refrain from drawing any strong conclusions from the late-time spectral evolution alone. On the other hand, the earlier epochs --- in particular, the presence of broad H$\alpha$ emission and the spectropolarimetric results of \cite{2014MNRAS.442.1166M} --- still argue for a true SN as the underlying event \citep{2013MNRAS.434.2721S}. Future late-time observations should be able to shed new light on the progenitor and fate of SN\,2009ip, as was done with photometry at 1--6000 days of SN impostor SN\,1997bs by \cite{2015MNRAS.452.2195A}, who suggest that its late-time data are more consistent with a genuine SN.

\section{Conclusions}\label{sec:conc}

In this work, we have presented and analyzed late-time optical photometric and spectroscopic observations of the ambiguous transient SN\,2009ip that we obtained with LCO and the Keck Observatory in 2013, 2014, and 2015. We showed that the light curve continued its steady linear decline throughout 2014 with a linear slope of $0.0030\pm0.0005$ $\rm mag\ day^{-1}$ in the $r^{\prime}$ band, and that this is similar both to observations and models of SNe\,IIn. We discussed how the eventual passing of the faintest previous magnitudes reported by Thoene et al. (2015) supports the inference that this decline continued throughout 2015, but emphasised that it is not unique proof that SN\,2009ip was a terminal explosion.

We have shown that the late-time slowly evolving spectrum remains dominated by the signatures of CSM interaction, and exhibits emission lines from species that are commonly found in SN\,IIn spectra. By comparing the emission-line profiles of H$\beta$ and \ion{He}{I} $\lambda7065$, we established that there exist CSM zones with little to no He that extend to higher velocities. The H$\alpha$/H$\beta$ intensity ratio is very large at late times, indicating ``Case C" recombination in a medium that is optically thick to the Balmer lines, in agreement with models for late-time CSM interaction. We evaluated the rate of decline in the H$\alpha$ luminosity and found that it is similar to observations of other SNe\,IIn and to models of interacting SNe. The nonzero multicomponent velocity structures of the H$\alpha$ emission, which are consistent with a disc-like CSM as suggested by \cite{2014AJ....147...23L} and \cite{2014MNRAS.442.1166M}, persist to late times. We introduced a parameter to quantify the degree of asymmetry in the H$\alpha$ emission and evaluated its evolution at late times, finding that the red side exhibits an excess flux which generally decreases with time and that the wavelength of peak flux shows no monotonic evolution. 

We made a qualitative comparison of the early-time and late-time characteristics and evolution of SN\,2009ip to simulated light curves and spectra of CSM interaction events that were recently presented by \cite{2016MNRAS.458.2094D}. Models that contain a genuine SN as the underlying power source can better explain the observations of SN\,2009ip. We also compared the early-time and late-time observations of SN\,2009ip to recently published data for SNe\,IIn and SN impostors, both of which are CSM-interaction events that appear to be driven by terminal and nonterminal underlying explosions, respectively. The late-time behaviour of SN\,2009ip has similarities with both of these types of events, but is less consistent with the spectra of SN impostors; moreover, the early-time data continue to support a genuine SN explosion. Ultimately, we conclude that our late-time data alone do not provide any strong independent constraints on the true nature of the underlying explosion, which remains shrouded in mystery. 

\section*{Acknowledgements}

We thank Emily Levesque for helpful conversations and V. Ashely Villar for access to their SN\,2010da spectra.
We are grateful to our anonymous referee for a thoughtful and constructive review. 

The supernova research of A.V.F.'s group at U.C. Berkeley is supported by Gary \& Cynthia Bengier, the Richard \& Rhoda Goldman Fund, the Christopher R. Redlich Fund, the TABASGO Foundation, and NSF grant AST--1211916. Some of A.V.F.'s work was completed at the Aspen Center for Physics, which is supported by NSF grant PHY--1066293; he thanks the Center for its hospitality during the black holes workshop in June and July of 2016.

This research makes use of observations from the Las Cumbres Observatory (LCO) network and is supported by National Science Foundation (NSF) grant AST--1313484.

This research is based in part on observations made at the W.~M.\ Keck Observatory; we are grateful to the Keck staff for their assistance, and we extend special thanks to those of Hawaiian ancestry on whose sacred mountain we are privileged to be guests. The W.~M.\ Keck Observatory is operated as a scientific partnership among the California Institute of Technology, the University of California, and the National Aeronautics and Space Administration (NASA); it was made possible by the generous financial support of the W.~M.\ Keck Foundation. We thank Daniel Perley and Brad Cenko for the use of, and assistance with, their Keck LRIS imaging and spectroscopy reduction pipeline\footnote{http://www.astro.caltech.edu/$\sim$dperley/programs/lpipe.html}. We also thank WeiKang Zheng, Kelsey Clubb, Brad Tucker, Patrick Kelly, and Isaac Shivvers for their assistance with the observations.

This research has made use of the NASA/IPAC Extragalactic Database (NED) which is operated by the Jet Propulsion Laboratory, California Institute of Technology, under contract with NASA.

This work has made use of the Weizmann interactive supernova data repository \citep{2012PASP..124..668Y} at \url{http://wiserep.weizmann.ac.il}.


\bibliographystyle{apj}
\bibliography{apj-jour,myrefs}

\begin{thebibliography}{}
\expandafter\ifx\csname natexlab\endcsname\relax\def\natexlab#1{#1}\fi

\bibitem[{{Adams} \& {Kochanek}(2015)}]{2015MNRAS.452.2195A}
{Adams}, S.~M., \& {Kochanek}, C.~S. 2015, \mnras, 452, 2195

\bibitem[{{Aretxaga} {et~al.}(1999){Aretxaga}, {Benetti}, {Terlevich},
  {Fabian}, {Cappellaro}, {Turatto}, \& {della Valle}}]{1999MNRAS.309..343A}
{Aretxaga}, I., {Benetti}, S., {Terlevich}, R.~J., {et~al.} 1999, \mnras, 309,
  343

\bibitem[{{Berger} {et~al.}(2009){Berger}, {Foley}, \&
  {Ivans}}]{2009ATel.2184....1B}
{Berger}, E., {Foley}, R., \& {Ivans}, I. 2009, The Astronomer's Telegram,
  2184, 1

\bibitem[{{Binder} {et~al.}(2016){Binder}, {Williams}, {Kong}, {Gaetz},
  {Plucinsky}, {Skillman}, \& {Dolphin}}]{2016MNRAS.457.1636B}
{Binder}, B., {Williams}, B.~F., {Kong}, A.~K.~H., {et~al.} 2016, \mnras, 457,
  1636

\bibitem[{{Brown} {et~al.}(2013){Brown}, {Baliber}, {Bianco}, {Bowman},
  {Burleson}, {Conway}, {Crellin}, {Depagne}, {De Vera}, {Dilday}, {Dragomir},
  {Dubberley}, {Eastman}, {Elphick}, {Falarski}, {Foale}, {Ford}, {Fulton},
  {Garza}, {Gomez}, {Graham}, {Greene}, {Haldeman}, {Hawkins}, {Haworth},
  {Haynes}, {Hidas}, {Hjelstrom}, {Howell}, {Hygelund}, {Lister}, {Lobdill},
  {Martinez}, {Mullins}, {Norbury}, {Parrent}, {Paulson}, {Petry}, {Pickles},
  {Posner}, {Rosing}, {Ross}, {Sand}, {Saunders}, {Shobbrook}, {Shporer},
  {Street}, {Thomas}, {Tsapras}, {Tufts}, {Valenti}, {Vander Horst}, {Walker},
  {White}, \& {Willis}}]{2013PASP..125.1031B}
{Brown}, T.~M., {Baliber}, N., {Bianco}, F.~B., {et~al.} 2013, \pasp, 125, 1031

\bibitem[{{Chevalier} \& {Fransson}(1994)}]{1994ApJ...420..268C}
{Chevalier}, R.~A., \& {Fransson}, C. 1994, \apj, 420, 268

\bibitem[{{Dessart} {et~al.}(2015){Dessart}, {Audit}, \&
  {Hillier}}]{2015MNRAS.449.4304D}
{Dessart}, L., {Audit}, E., \& {Hillier}, D.~J. 2015, \mnras, 449, 4304

\bibitem[{{Dessart} {et~al.}(2016){Dessart}, {Hillier}, {Audit}, {Livne}, \&
  {Waldman}}]{2016MNRAS.458.2094D}
{Dessart}, L., {Hillier}, D.~J., {Audit}, E., {Livne}, E., \& {Waldman}, R.
  2016, \mnras, 458, 2094

\bibitem[{{Drake} {et~al.}(2010){Drake}, {Prieto}, {Djorgovski}, {Mahabal},
  {Graham}, {Williams}, {McNaught}, {Catelan}, {Christensen}, {Beshore},
  {Larson}, \& {Howerton}}]{2010ATel.2897....1D}
{Drake}, A.~J., {Prieto}, J.~L., {Djorgovski}, S.~G., {et~al.} 2010, The
  Astronomer's Telegram, 2897, 1

\bibitem[{{Elias-Rosa} {et~al.}(2016){Elias-Rosa}, {Pastorello}, {Benetti},
  {Cappellaro}, {Taubenberger}, {Terreran}, {Fraser}, {Brown}, {Tartaglia},
  {Morales-Garoffolo}, {Harmanen}, {Richardson}, {Artigau}, {Tomasella},
  {Margutti}, {Smartt}, {Dennefeld}, {Turatto}, {Anupama}, {Arbour}, {Berton},
  {Bjorkman}, {Boles}, {Briganti}, {Chornock}, {Ciabattari}, {Cortini},
  {Dimai}, {Gerhartz}, {Itagaki}, {Kotak}, {Mancini}, {Martinelli},
  {Milisavljevic}, {Misra}, {Ochner}, {Patnaude}, {Polshaw}, {Sahu}, \&
  {Zaggia}}]{2016MNRAS.463.3894E}
{Elias-Rosa}, N., {Pastorello}, A., {Benetti}, S., {et~al.} 2016, \mnras, 463,
  3894

\bibitem[{{Faber} {et~al.}(2003){Faber}, {Phillips}, {Kibrick}, {Alcott},
  {Allen}, {Burrous}, {Cantrall}, {Clarke}, {Coil}, {Cowley}, {Davis}, {Deich},
  {Dietsch}, {Gilmore}, {Harper}, {Hilyard}, {Lewis}, {McVeigh}, {Newman},
  {Osborne}, {Schiavon}, {Stover}, {Tucker}, {Wallace}, {Wei}, {Wirth}, \&
  {Wright}}]{2003SPIE.4841.1657F}
{Faber}, S.~M., {Phillips}, A.~C., {Kibrick}, R.~I., {et~al.} 2003, in Society
  of Photo-Optical Instrumentation Engineers (SPIE) Conference Series, Vol.
  4841, Instrument Design and Performance for Optical/Infrared Ground-based
  Telescopes, ed. M.~{Iye} \& A.~F.~M. {Moorwood}, 1657--1669

\bibitem[{{Fassia} {et~al.}(2001){Fassia}, {Meikle}, {Chugai}, {Geballe},
  {Lundqvist}, {Walton}, {Pollacco}, {Veilleux}, {Wright}, {Pettini}, {Kerr},
  {Puchnarewicz}, {Puxley}, {Irwin}, {Packham}, {Smartt}, \&
  {Harmer}}]{2001MNRAS.325..907F}
{Fassia}, A., {Meikle}, W.~P.~S., {Chugai}, N., {et~al.} 2001, \mnras, 325, 907

\bibitem[{{Filippenko}(1982)}]{1982PASP...94..715F}
{Filippenko}, A.~V. 1982, \pasp, 94, 715

\bibitem[{{Foley} {et~al.}(2011){Foley}, {Berger}, {Fox}, {Levesque},
  {Challis}, {Ivans}, {Rhoads}, \& {Soderberg}}]{2011ApJ...732...32F}
{Foley}, R.~J., {Berger}, E., {Fox}, O., {et~al.} 2011, \apj, 732, 32

\bibitem[{{Fox} {et~al.}(2015){Fox}, {Silverman}, {Filippenko}, {Mauerhan},
  {Becker}, {Borish}, {Cenko}, {Clubb}, {Graham}, {Hsiao}, {Kelly}, {Lee},
  {Marion}, {Milisavljevic}, {Parrent}, {Shivvers}, {Skrutskie}, {Smith},
  {Wilson}, \& {Zheng}}]{2015MNRAS.447..772F}
{Fox}, O.~D., {Silverman}, J.~M., {Filippenko}, A.~V., {et~al.} 2015, \mnras,
  447, 772

\bibitem[{{Fraser} {et~al.}(2013){Fraser}, {Inserra}, {Jerkstrand}, {Kotak},
  {Pignata}, {Benetti}, {Botticella}, {Bufano}, {Childress}, {Mattila},
  {Pastorello}, {Smartt}, {Turatto}, {Yuan}, {Anderson}, {Bayliss}, {Bauer},
  {Chen}, {F{\"o}rster Bur{\'o}n}, {Gal-Yam}, {Haislip}, {Knapic}, {Le
  Guillou}, {Marchi}, {Mazzali}, {Molinaro}, {Moore}, {Reichart}, {Smareglia},
  {Smith}, {Sternberg}, {Sullivan}, {Tak{\'a}ts}, {Tucker}, {Valenti}, {Yaron},
  {Young}, \& {Zhou}}]{2013MNRAS.433.1312F}
{Fraser}, M., {Inserra}, C., {Jerkstrand}, A., {et~al.} 2013, \mnras, 433, 1312

\bibitem[{{Fraser} {et~al.}(2015){Fraser}, {Kotak}, {Pastorello}, {Jerkstrand},
  {Smartt}, {Chen}, {Childress}, {Gilmore}, {Inserra}, {Kankare}, {Margheim},
  {Mattila}, {Valenti}, {Ashall}, {Benetti}, {Botticella}, {Bauer}, {Campbell},
  {Elias-Rosa}, {Fleury}, {Gal-Yam}, {Hachinger}, {Howell}, {Le Guillou},
  {L{\'e}get}, {Morales-Garoffolo}, {Polshaw}, {Spiro}, {Sullivan},
  {Taubenberger}, {Turatto}, {Walker}, {Young}, \&
  {Zhang}}]{2015MNRAS.453.3886F}
{Fraser}, M., {Kotak}, R., {Pastorello}, A., {et~al.} 2015, \mnras, 453, 3886

\bibitem[{{Goranskij} {et~al.}(2016){Goranskij}, {Barsukova}, {Valeev},
  {Tsvetkov}, {Volkov}, {Metlov}, \& {Zharova}}]{2016arXiv160900731G}
{Goranskij}, V.~P., {Barsukova}, E.~A., {Valeev}, A.~F., {et~al.} 2016, ArXiv
  e-prints, arXiv:1609.00731

\bibitem[{{Graham} {et~al.}(2014){Graham}, {Sand}, {Valenti}, {Howell},
  {Parrent}, {Halford}, {Zaritsky}, {Bianco}, {Rest}, \&
  {Dilday}}]{2014ApJ...787..163G}
{Graham}, M.~L., {Sand}, D.~J., {Valenti}, S., {et~al.} 2014, \apj, 787, 163

\bibitem[{{Grandi}(1980)}]{1980ApJ...238...10G}
{Grandi}, S.~A. 1980, \apj, 238, 10

\bibitem[{{Jencson} {et~al.}(2016){Jencson}, {Prieto}, {Kochanek}, {Shappee},
  {Stanek}, \& {Pogge}}]{2016MNRAS.456.2622J}
{Jencson}, J.~E., {Prieto}, J.~L., {Kochanek}, C.~S., {et~al.} 2016, \mnras,
  456, 2622

\bibitem[{{Jerkstrand} {et~al.}(2014){Jerkstrand}, {Smartt}, {Fraser},
  {Fransson}, {Sollerman}, {Taddia}, \& {Kotak}}]{2014MNRAS.439.3694J}
{Jerkstrand}, A., {Smartt}, S.~J., {Fraser}, M., {et~al.} 2014, \mnras, 439,
  3694

\bibitem[{Kramida {et~al.}(2015)Kramida, {Yu.~Ralchenko}, Reader, \& {and NIST
  ASD Team}}]{NIST_ASD}
Kramida, A., {Yu.~Ralchenko}, Reader, J., \& {and NIST ASD Team}. 2015, {NIST
  Atomic Spectra Database (ver. 5.3), [Online]. Available:
  {\tt{http://physics.nist.gov/asd}} [2015, December 5]. National Institute of
  Standards and Technology Gaithersburg, MD.}

\bibitem[{{Landt} {et~al.}(2008){Landt}, {Bentz}, {Ward}, {Elvis}, {Peterson},
  {Korista}, \& {Karovska}}]{2008ApJS..174..282L}
{Landt}, H., {Bentz}, M.~C., {Ward}, M.~J., {et~al.} 2008, \apjs, 174, 282

\bibitem[{{Levesque} {et~al.}(2014){Levesque}, {Stringfellow}, {Ginsburg},
  {Bally}, \& {Keeney}}]{2014AJ....147...23L}
{Levesque}, E.~M., {Stringfellow}, G.~S., {Ginsburg}, A.~G., {Bally}, J., \&
  {Keeney}, B.~A. 2014, \aj, 147, 23

\bibitem[{{Li} {et~al.}(2009){Li}, {Smith}, {Miller}, \&
  {Filippenko}}]{2009ATel.2212....1L}
{Li}, W., {Smith}, N., {Miller}, A.~A., \& {Filippenko}, A.~V. 2009, The
  Astronomer's Telegram, 2212, 1

\bibitem[{{Margutti} {et~al.}(2014){Margutti}, {Milisavljevic}, {Soderberg},
  {Chornock}, {Zauderer}, {Murase}, {Guidorzi}, {Sanders}, {Kuin}, {Fransson},
  {Levesque}, {Chandra}, {Berger}, {Bianco}, {Brown}, {Challis},
  {Chatzopoulos}, {Cheung}, {Choi}, {Chomiuk}, {Chugai}, {Contreras}, {Drout},
  {Fesen}, {Foley}, {Fong}, {Friedman}, {Gall}, {Gehrels}, {Hjorth}, {Hsiao},
  {Kirshner}, {Im}, {Leloudas}, {Lunnan}, {Marion}, {Martin}, {Morrell},
  {Neugent}, {Omodei}, {Phillips}, {Rest}, {Silverman}, {Strader},
  {Stritzinger}, {Szalai}, {Utterback}, {Vinko}, {Wheeler}, {Arnett},
  {Campana}, {Chevalier}, {Ginsburg}, {Kamble}, {Roming}, {Pritchard}, \&
  {Stringfellow}}]{2014ApJ...780...21M}
{Margutti}, R., {Milisavljevic}, D., {Soderberg}, A.~M., {et~al.} 2014, \apj,
  780, 21

\bibitem[{{Marziani} {et~al.}(2014){Marziani}, {Mart{\'{\i}}nez-Aldama},
  {Dultzin}, \& {Sulentic}}]{2014AstRv...9a..29M}
{Marziani}, P., {Mart{\'{\i}}nez-Aldama}, M.~L., {Dultzin}, D., \& {Sulentic},
  J.~W. 2014, The Astronomical Review, 9, 29

\bibitem[{{Mauerhan} \& {Smith}(2012)}]{2012MNRAS.424.2659M}
{Mauerhan}, J., \& {Smith}, N. 2012, \mnras, 424, 2659

\bibitem[{{Mauerhan} {et~al.}(2014){Mauerhan}, {Williams}, {Smith}, {Smith},
  {Filippenko}, {Hoffman}, {Milne}, {Leonard}, {Clubb}, {Fox}, \&
  {Kelly}}]{2014MNRAS.442.1166M}
{Mauerhan}, J., {Williams}, G.~G., {Smith}, N., {et~al.} 2014, \mnras, 442,
  1166

\bibitem[{{Mauerhan} {et~al.}(2013){Mauerhan}, {Smith}, {Filippenko},
  {Blanchard}, {Blanchard}, {Casper}, {Cenko}, {Clubb}, {Cohen}, {Fuller},
  {Li}, \& {Silverman}}]{2013MNRAS.430.1801M}
{Mauerhan}, J.~C., {Smith}, N., {Filippenko}, A.~V., {et~al.} 2013, \mnras,
  430, 1801

\bibitem[{{Maza} {et~al.}(2009){Maza}, {Hamuy}, {Antezana}, {Gonzalez},
  {Lopez}, {Silva}, {Folatelli}, {Iturra}, {Cartier}, {Forster}, {Marchi},
  {Rojas}, {Pignata}, {Conuel}, {Reichart}, {Ivarsen}, {Haislip}, {Crain},
  {Foster}, {Nysewander}, \& {Lacluyze}}]{2009CBET.1928....1M}
{Maza}, J., {Hamuy}, M., {Antezana}, R., {et~al.} 2009, Central Bureau
  Electronic Telegrams, 1928, 1

\bibitem[{{Miller} {et~al.}(2009){Miller}, {Li}, {Nugent}, {Bloom},
  {Filippenko}, \& {Merritt}}]{2009ATel.2183....1M}
{Miller}, A.~A., {Li}, W., {Nugent}, P.~E., {et~al.} 2009, The Astronomer's
  Telegram, 2183, 1

\bibitem[{{Moriya}(2015)}]{2015ApJ...803L..26M}
{Moriya}, T.~J. 2015, \apjl, 803, L26

\bibitem[{{Ofek} {et~al.}(2013){Ofek}, {Lin}, {Kouveliotou}, {Younes},
  {G{\"o}{\v g}{\"u}{\c s}}, {Kasliwal}, \& {Cao}}]{2013ApJ...768...47O}
{Ofek}, E.~O., {Lin}, L., {Kouveliotou}, C., {et~al.} 2013, \apj, 768, 47

\bibitem[{{Ofek} {et~al.}(2014){Ofek}, {Sullivan}, {Shaviv}, {Steinbok},
  {Arcavi}, {Gal-Yam}, {Tal}, {Kulkarni}, {Nugent}, {Ben-Ami}, {Kasliwal},
  {Cenko}, {Laher}, {Surace}, {Bloom}, {Filippenko}, {Silverman}, \&
  {Yaron}}]{2014ApJ...789..104O}
{Ofek}, E.~O., {Sullivan}, M., {Shaviv}, N.~J., {et~al.} 2014, \apj, 789, 104

\bibitem[{{Oke} {et~al.}(1995){Oke}, {Cohen}, {Carr}, {Cromer}, {Dingizian},
  {Harris}, {Labrecque}, {Lucinio}, {Schaal}, {Epps}, \&
  {Miller}}]{1995PASP..107..375O}
{Oke}, J.~B., {Cohen}, J.~G., {Carr}, M., {et~al.} 1995, \pasp, 107, 375

\bibitem[{{Pastorello} {et~al.}(2013){Pastorello}, {Cappellaro}, {Inserra},
  {Smartt}, {Pignata}, {Benetti}, {Valenti}, {Fraser}, {Tak{\'a}ts}, {Benitez},
  {Botticella}, {Brimacombe}, {Bufano}, {Cellier-Holzem}, {Costado}, {Cupani},
  {Curtis}, {Elias-Rosa}, {Ergon}, {Fynbo}, {Hambsch}, {Hamuy}, {Harutyunyan},
  {Ivarson}, {Kankare}, {Martin}, {Kotak}, {LaCluyze}, {Maguire}, {Mattila},
  {Maza}, {McCrum}, {Miluzio}, {Norgaard-Nielsen}, {Nysewander}, {Ochner},
  {Pan}, {Pumo}, {Reichart}, {Tan}, {Taubenberger}, {Tomasella}, {Turatto}, \&
  {Wright}}]{2013ApJ...767....1P}
{Pastorello}, A., {Cappellaro}, E., {Inserra}, C., {et~al.} 2013, \apj, 767, 1

\bibitem[{{Patat} {et~al.}(2011){Patat}, {Taubenberger}, {Benetti},
  {Pastorello}, \& {Harutyunyan}}]{2011A&A...527L...6P}
{Patat}, F., {Taubenberger}, S., {Benetti}, S., {Pastorello}, A., \&
  {Harutyunyan}, A. 2011, \aap, 527, L6

\bibitem[{{Pozzo} {et~al.}(2004){Pozzo}, {Meikle}, {Fassia}, {Geballe},
  {Lundqvist}, {Chugai}, \& {Sollerman}}]{2004MNRAS.352..457P}
{Pozzo}, M., {Meikle}, W.~P.~S., {Fassia}, A., {et~al.} 2004, \mnras, 352, 457

\bibitem[{{Prieto} {et~al.}(2013){Prieto}, {Brimacombe}, {Drake}, \&
  {Howerton}}]{2013ApJ...763L..27P}
{Prieto}, J.~L., {Brimacombe}, J., {Drake}, A.~J., \& {Howerton}, S. 2013,
  \apjl, 763, L27

\bibitem[{{Rockosi} {et~al.}(2010){Rockosi}, {Stover}, {Kibrick}, {Lockwood},
  {Peck}, {Cowley}, {Bolte}, {Adkins}, {Alcott}, {Allen}, {Brown}, {Cabak},
  {Deich}, {Hilyard}, {Kassis}, {Lanclos}, {Lewis}, {Pfister}, {Phillips},
  {Robinson}, {Saylor}, {Thompson}, {Ward}, {Wei}, \&
  {Wright}}]{2010SPIE.7735E..0RR}
{Rockosi}, C., {Stover}, R., {Kibrick}, R., {et~al.} 2010, in \procspie, Vol.
  7735, Ground-based and Airborne Instrumentation for Astronomy III, 77350R

\bibitem[{{Smith} {et~al.}(2016){Smith}, {Andrews}, \&
  {Mauerhan}}]{2016arXiv160701056S}
{Smith}, N., {Andrews}, J.~E., \& {Mauerhan}, J.~C. 2016, ArXiv e-prints,
  arXiv:1607.01056

\bibitem[{{Smith} {et~al.}(2008){Smith}, {Chornock}, {Li}, {Ganeshalingam},
  {Silverman}, {Foley}, {Filippenko}, \& {Barth}}]{2008ApJ...686..467S}
{Smith}, N., {Chornock}, R., {Li}, W., {et~al.} 2008, \apj, 686, 467

\bibitem[{{Smith} \& {Mauerhan}(2012)}]{2012ATel.4412....1S}
{Smith}, N., \& {Mauerhan}, J. 2012, The Astronomer's Telegram, 4412, 1

\bibitem[{{Smith} {et~al.}(2013){Smith}, {Mauerhan}, {Kasliwal}, \&
  {Burgasser}}]{2013MNRAS.434.2721S}
{Smith}, N., {Mauerhan}, J.~C., {Kasliwal}, M.~M., \& {Burgasser}, A.~J. 2013,
  \mnras, 434, 2721

\bibitem[{{Smith} {et~al.}(2014){Smith}, {Mauerhan}, \&
  {Prieto}}]{2014MNRAS.438.1191S}
{Smith}, N., {Mauerhan}, J.~C., \& {Prieto}, J.~L. 2014, \mnras, 438, 1191

\bibitem[{{Smith} {et~al.}(2010){Smith}, {Miller}, {Li}, {Filippenko},
  {Silverman}, {Howard}, {Nugent}, {Marcy}, {Bloom}, {Ghez}, {Lu}, {Yelda},
  {Bernstein}, \& {Colucci}}]{2010AJ....139.1451S}
{Smith}, N., {Miller}, A., {Li}, W., {et~al.} 2010, \aj, 139, 1451

\bibitem[{{Smith} {et~al.}(2017){Smith}, {Kilpatrick}, {Mauerhan}, {Andrews},
  {Margutti}, {Fong}, {Graham}, {Zheng}, {Kelly}, {Filippenko}, \&
  {Fox}}]{2017MNRAS.466.3021S}
{Smith}, N., {Kilpatrick}, C.~D., {Mauerhan}, J.~C., {et~al.} 2017, \mnras,
  466, 3021

\bibitem[{{Sollerman} {et~al.}(1998){Sollerman}, {Cumming}, \&
  {Lundqvist}}]{1998ApJ...493..933S}
{Sollerman}, J., {Cumming}, R.~J., \& {Lundqvist}, P. 1998, \apj, 493, 933

\bibitem[{{Stritzinger} {et~al.}(2012){Stritzinger}, {Taddia}, {Fransson},
  {Fox}, {Morrell}, {Phillips}, {Sollerman}, {Anderson}, {Boldt}, {Brown},
  {Campillay}, {Castellon}, {Contreras}, {Folatelli}, {Habergham}, {Hamuy},
  {Hjorth}, {James}, {Krzeminski}, {Mattila}, {Persson}, \&
  {Roth}}]{2012ApJ...756..173S}
{Stritzinger}, M., {Taddia}, F., {Fransson}, C., {et~al.} 2012, \apj, 756, 173

\bibitem[{{Tartaglia} {et~al.}(2015){Tartaglia}, {Pastorello}, {Taubenberger},
  {Cappellaro}, {Maund}, {Benetti}, {Boles}, {Bufano}, {Duszanowicz},
  {Elias-Rosa}, {Harutyunyan}, {Hermansson}, {H{\"o}flich}, {Maguire},
  {Navasardyan}, {Smartt}, {Taddia}, \& {Turatto}}]{2015MNRAS.447..117T}
{Tartaglia}, L., {Pastorello}, A., {Taubenberger}, S., {et~al.} 2015, \mnras,
  447, 117

\bibitem[{{Tartaglia} {et~al.}(2016{\natexlab{a}}){Tartaglia}, {Pastorello},
  {Sullivan}, {Baltay}, {Rabinowitz}, {Nugent}, {Drake}, {Djorgovski},
  {Gal-Yam}, {Fabrika}, {Barsukova}, {Goranskij}, {Valeev}, {Fatkhullin},
  {Schulze}, {Mehner}, {Bauer}, {Taubenberger}, {Nordin}, {Valenti}, {Howell},
  {Benetti}, {Cappellaro}, {Fasano}, {Elias-Rosa}, {Barbieri}, {Bettoni},
  {Harutyunyan}, {Kangas}, {Kankare}, {Martin}, {Mattila}, {Morales-Garoffolo},
  {Ochner}, {Rebbapragada}, {Terreran}, {Tomasella}, {Turatto}, {Verroi}, \&
  {Wo{\'z}niak}}]{2016MNRAS.459.1039T}
{Tartaglia}, L., {Pastorello}, A., {Sullivan}, M., {et~al.} 2016{\natexlab{a}},
  \mnras, 459, 1039

\bibitem[{{Tartaglia} {et~al.}(2016{\natexlab{b}}){Tartaglia}, {Elias-Rosa},
  {Pastorello}, {Benetti}, {Taubenberger}, {Cappellaro}, {Cortini}, {Granata},
  {Ishida}, {Morales-Garoffolo}, {Noebauer}, {Ochner}, {Tomasella}, \&
  {Zaggia}}]{2016ApJ...823L..23T}
{Tartaglia}, L., {Elias-Rosa}, N., {Pastorello}, A., {et~al.}
  2016{\natexlab{b}}, \apjl, 823, L23

\bibitem[{{Thoene} {et~al.}(2015){Thoene}, {de Ugarte Postigo}, {Leloudas},
  {Cano}, \& {Maeda}}]{2015ATel.8417....1T}
{Thoene}, C., {de Ugarte Postigo}, A., {Leloudas}, G., {Cano}, Z., \& {Maeda},
  K. 2015, The Astronomer's Telegram, 8417

\bibitem[{{Valenti} {et~al.}(2016){Valenti}, {Howell}, {Stritzinger}, {Graham},
  {Hosseinzadeh}, {Arcavi}, {Bildsten}, {Jerkstrand}, {McCully}, {Pastorello},
  {Piro}, {Sand}, {Smartt}, {Terreran}, {Baltay}, {Benetti}, {Brown},
  {Filippenko}, {Fraser}, {Rabinowitz}, {Sullivan}, \&
  {Yuan}}]{2016MNRAS.459.3939V}
{Valenti}, S., {Howell}, D.~A., {Stritzinger}, M.~D., {et~al.} 2016, \mnras,
  459, 3939

\bibitem[{{Van Dyk} {et~al.}(2000){Van Dyk}, {Peng}, {King}, {Filippenko},
  {Treffers}, {Li}, \& {Richmond}}]{2000PASP..112.1532V}
{Van Dyk}, S.~D., {Peng}, C.~Y., {King}, J.~Y., {et~al.} 2000, \pasp, 112, 1532

\bibitem[{{Villar} {et~al.}(2016){Villar}, {Berger}, {Chornock}, {Margutti},
  {Laskar}, {Brown}, {Blanchard}, {Czekala}, {Lunnan}, \&
  {Reynolds}}]{2016ApJ...830...11V}
{Villar}, V.~A., {Berger}, E., {Chornock}, R., {et~al.} 2016, \apj, 830, 11

\bibitem[{{Weiler} {et~al.}(1998){Weiler}, {Van Dyk}, {Montes}, {Panagia}, \&
  {Sramek}}]{1998ApJ...500...51W}
{Weiler}, K.~W., {Van Dyk}, S.~D., {Montes}, M.~J., {Panagia}, N., \& {Sramek},
  R.~A. 1998, \apj, 500, 51

\bibitem[{{Xu} {et~al.}(1992){Xu}, {McCray}, {Oliva}, \&
  {Randich}}]{1992ApJ...386..181X}
{Xu}, Y., {McCray}, R., {Oliva}, E., \& {Randich}, S. 1992, \apj, 386, 181

\bibitem[{{Yaron} \& {Gal-Yam}(2012)}]{2012PASP..124..668Y}
{Yaron}, O., \& {Gal-Yam}, A. 2012, \pasp, 124, 668

\bibitem[{{Zhang} {et~al.}(2012){Zhang}, {Wang}, {Wu}, {Chen}, {Chen}, {Liu},
  {Huang}, {Liang}, {Zhao}, {Lin}, {Wang}, {Dennefeld}, {Zhang}, {Zhai}, {Wu},
  {Fan}, {Zou}, {Zhou}, \& {Ma}}]{2012AJ....144..131Z}
{Zhang}, T., {Wang}, X., {Wu}, C., {et~al.} 2012, \aj, 144, 131

\end{thebibliography}

\label{lastpage}

\end{document}